\documentclass[11pt]{elsarticle}

\makeatletter
  \long\def\pprintMaketitle{\clearpage
  \iflongmktitle\if@twocolumn\let\columnwidth=\textwidth\fi\fi
  \resetTitleCounters
  \def\baselinestretch{1}%
  \printFirstPageNotes
  \begin{center}%
 \thispagestyle{pprintTitle}%
   \def\baselinestretch{1}%
    {\large\bf\@title}\par\vskip5pt
    \normalsize\elsauthors\par\vskip5pt
    \footnotesize\itshape\elsaddress\par\vskip10pt
    \end{center}%
  \gdef\thefootnote{\arabic{footnote}}%
  }
\makeatother

\usepackage{lmodern}
\usepackage[utf8]{inputenc}
\usepackage[T1]{fontenc}
\usepackage{amsmath}
\usepackage{amsfonts}
\usepackage{lipsum}
\usepackage{graphicx}
\usepackage{xspace}
\usepackage{epsf}
\usepackage{setspace}
\usepackage{float}
\usepackage{soul}

\usepackage{mwe}
\usepackage{float}
\usepackage{subfig}
\usepackage{graphicx}
\usepackage{epstopdf}
\usepackage{caption}
\usepackage[inkscapelatex=false]{svg}
\usepackage{placeins}
\usepackage{algorithm}
\usepackage{algpseudocode}
\usepackage{comment}
\usepackage{amsmath}
\usepackage{bm}
\usepackage[normalem]{ulem}
\usepackage[
    protrusion=true,
    activate={true,nocompatibility},
    final,
    tracking=true,
    kerning=true,
    spacing=true,
    factor=1100]{microtype}
\SetTracking{encoding={*}, shape=sc}{40}
\microtypecontext{spacing=nonfrench}

\usepackage[margin=1in]{geometry}
\usepackage{abstract}
\usepackage{booktabs}

\usepackage{hyperref}

\usepackage{graphicx}

\usepackage{stmaryrd}
\SetSymbolFont{stmry}{bold}{U}{stmry}{m}{n}

\usepackage{placeins}
\usepackage{mathrsfs}

 \usepackage{bm}
  \usepackage{subfig}
  \usepackage{caption}
\usepackage{algorithm}
\usepackage{algpseudocode}
\usepackage{url}
\usepackage[parfill]{parskip}

   \usepackage{graphicx}
 \usepackage{float}
\usepackage{epstopdf, epsfig}
\usepackage{physics}

\usepackage[]{xcolor}
\usepackage{stmaryrd}
\usepackage{graphics}
\usepackage{appendix}

\usepackage{kantlipsum}

\makeatletter
\renewcommand{\Function}[2]{%
  \csname ALG@cmd@\ALG@L @Function\endcsname{#1}{#2}%
  \def\jayden@currentfunction{#1}%
}
\newcommand{\funclabel}[1]{%
  \@bsphack
  \protected@write\@auxout{}{%
    \string\newlabel{#1}{{\jayden@currentfunction}{\thepage}}%
  }%
  \@esphack
}

 \usepackage{amsmath,amssymb}

\begin{document}

\hypersetup{
  pdfauthor=author,
}

\begin{frontmatter}
\title{Lyapunov stability analysis of the chaotic flow past two square cylinders}


\author[add1]{Sidhartha Sahu}
\author[add1]{George Papadakis \corref{cor1}} \ead{g.papadakis@imperial.ac.uk}

\address[add1]{Department of Aeronautics, Imperial College London, Exhibition Rd, London SW7 2AZ, UK}
\cortext[cor1]{Corresponding author}
    \date{}

\end{frontmatter}

\begin{abstract}

We investigate the stability of the flow past two side-by-side square cylinders (at Reynolds number 200 and gap ratio 1) using tools from dynamical systems theory. The flow is highly irregular due to the complex interaction between the flapping jet emanating from the gap and the vortices shed in the wake. We first perform Spectral Proper Orthogonal Decomposition (SPOD) to understand the flow characteristics. We then conduct Lyapunov stability analysis by linearizing the Navier-Stokes equations around the irregular base flow and find that it has two positive Lyapunov exponents. The Covariant Lyapunov Vectors (CLVs) are also computed. Contours of the time-averaged CLVs reveal that the footprint of the leading CLV is in the near-wake, whereas the other CLVs peak further downstream, indicating distinct regions of instability. SPOD of the two unstable CLVs is then employed to extract the  dominant coherent structures and oscillation frequencies in the tangent space. For the leading CLV, the two dominant frequencies match closely with the prevalent frequencies in the drag coefficient spectrum, and correspond to instabilities due to vortex shedding and jet-flapping. The second unstable CLV captures the subharmonic instability of the shedding frequency. Global linear stability analysis (GLSA) of the time-averaged flow identifies a neutral eigenmode that resembles the leading SPOD mode of the first CLV, with a very similar structure and frequency. However, while GLSA predicts neutrality, Lyapunov analysis reveals that this direction is unstable, exposing the inherent limitations of the GLSA when applied to chaotic flows.
\end{abstract}


\section{Introduction}
Flows around cylindrical engineering structures are ubiquitous, and understanding their stability characteristics is important. For example, in urban settings, understanding of the flow stability past square-shaped buildings is essential for maintaining structural integrity and minimizing wind-induced vibrations \citep{tall_buildings}. Likewise, vortex-induced motions in offshore structures such as semi-submersible platforms, driven by vortex shedding and wake dynamics, can lead to pronounced transverse and yaw oscillations that amplify structural responses \citep{foursquares}. Stability analysis of these flows can provide critical insights and help with mitigating risks posed by oscillatory phenomena.

Often these cylindrical structures are arranged in side-by-side configurations, resembling two-dimensional flows around adjacent square cylinders. This arrangement exhibits a rich tapestry of flow patterns. For example, at  Reynolds number $4.7 \times 10^4$ \cite{Alam_Zhou_and_Wang_2011} identified 4 regimes (single-body, two-frequency, transition, and coupled vortex street) depending on the gap ratio. \citet{ma_wake_2017} considered lower $Re$ numbers (between $16-200$) and gap ratios $0-10D$ (where $D$ is the edge length of the square cylinders) and identified 9 flow regimes. For a range of gap ratios (that depends on the Reynolds number) the vortex streets behind the two cylinders are asymmetric causing the jet emanating from the gap  to flap up and down. This behavior is also well documented for circular cylinder flows and is known as “flip-flopping” or “jet switching” \citep{Kim_Durbin_1988_flipflop,Moretti_jet_switching1993}. As the gap increases, the system transitions to synchronized, symmetric, or anti-symmetric vortex shedding regimes, see \cite{twosquarecylinders_latticeboltzman_Rao2008, ma_wake_2017}. High-fidelity simulations have captured these intricate wake interference patterns \citep{Extreme_events_Zhou_Nagata_Sakai_Watanabe_2019, Energy_transfer_two_squares_Zhou2020}. 

Global linear stability analysis (GLSA) offers a natural starting point for understanding the origin of some of these patterns. At its simplest form, GLSA considers a given steady base flow and investigates how small disturbances evolve when superimposed on the base flow \citep{Theofilis_2011, schmid_nonmodal_2007}. This involves the solution of an eigenvalue problem which is solved iteratively using Krylov subspace methods \citep{edwards_krylov_1994}. If the perturbations grow exponentially in time, then a primary instability has been detected. The primary instability of the flow past a square cylinder was considered by \cite{Yoon_et_al_2010,Park_Yang_2016,Jiang_Cheng_An_2018,GLSA_rectangles_JFM2021_Chiarini_Quadrio_Auteri}. It was found that its onset is similar to that of a circular cylinder with the flow undergoing a Hopf bifurcation. The most recent GLSA of  \cite{GLSA_rectangles_JFM2021_Chiarini_Quadrio_Auteri} reports a critical $Re_c=44.56$, which is slightly smaller than the $Re_c=46.6$ of a circular cylinder.  

For $Re>Re_c$, the flow becomes periodic and at a certain $Re$ a secondary instability may develop. This is captured by linearising around the two-dimensional periodic base flow and evaluating the evolution of three-dimensional perturbations superimposed on this base flow; this is known as Floquet stability analysis, \cite{Davis_1976}. The solution of the linearised equations can be decomposed into a sum of terms of the form $\tilde{\boldsymbol{u}}(x,y,z,t)e^{\sigma t}$ where $\tilde{\boldsymbol{u}}(x,y,z,t)$ is also periodic (with same period as the base flow) and $\sigma$ is the Floquet exponent. \cite{Barkley_Henderson_1996} reported the first three-dimensional secondary  stability analysis for the flow around a circular cylinder. For a square cylinder, Floquet analysis has been carried out by \cite{Yoon_et_al_2010, Park_Yang_2016, Jiang_Cheng_An_2018}. \cite{Carini_Giannetti_Auteri_2014} found that the ``jet switching" or ``flip-flopping" that has been reported by many investigators arises as a secondary instability of the periodic in-phase synchronised shedding over two side-by-side cylinders. 

Further increase in Reynolds number results in highly irregular flows and Floquet analysis is no longer valid. In such cases, a popular approach is GLSA around the time-average flow. This flow however satisfies the Reynolds-averaged Navier-Stokes (RANS) equations that contain Reynolds stresses. When the RANS equations are linearised to perform GLSA, the variation of the Reynolds stresses is (usually) neglected. In order to take this variation into account, a turbulence model is required, see \cite{Reynolds_Hussain_1972} . Either way, GLSA around a time-average flow has limitations. However, under specific circumstances, useful information can be obtained. For example, when the time-averaged wake behind a circular cylinder is used as the base flow, the frequency of the most unstable eigenmonde matches well the experimentally observed frequency and the growth rate is predicted to have a very small value, close to 0 \citep{pier_2002, Barkley_2006}. \cite{Sipp_lebedev_2007} gave a theoretical proof of this result by performing a weakly non-linear analysis which is valid close to $Re_c$. They calculated the constants of the Stuart-Landau amplitude equations and established conditions under which the linear analysis of the mean flow will yield the correct non-linear frequency of the limit cycle. They showed that the conditions were satisfied by the cylinder flow, but not for an open cavity flow. 
More general theoretical conditions for the use and meaning of a stability analysis around a mean flow are provided by 
\citet{Beneddine_Sipp_Arnault_Dandois_Lesshafft_2016}.In summary, while GLSA around a time-average flow can correctly identify and characterize instabilities in some irregular (even turbulent) flows, it should be applied with caution. A more general framework is therefore necessary to rigorously assess flow stability for complex, unsteady, laminar or turbulent flows.

Lyapunov stability analysis offers such a mathematically rigorous framework. It is based on linearisation around a general unsteady base flow and provides Lyapunov exponents (LEs) and Covariant Lyapunov Vectors (CLVs), that generalize the eigenvalues and eigenmodes of a traditional GLSA (or Floquet) analysis, respectively. Indeed,  \cite{Trevisan_Pancotti_1998} have shown that CLVs coincide with the Floquet modes and the LEs with the Floquet exponents for time-periodic base flows. Lyapunov stability analysis has been recognized as a tool for characterizing chaotic dynamics since the late 1970s, when effective algorithms were independently proposed by \cite{Shimada1979} and \cite{benettin_lyapunov_1980} to calculate LEs. The latter are independent of the norm used to compute them, and can be employed to characterise key physical properties, such as sensitivity to initial conditions, dynamical entropies, and fractal dimensions such as the Kaplan-Yorke dimension \citep{Eckmann_Ruelle_1985}. With respect to the tangent space, Gram–Schmidt (GS) vectors (which arise directly from the algorithm used to compute LEs) offer limited interpretability, since they are orthogonal by construction even at points on the attractor where the stable and unstable subspaces are nearly tangent. On the other hand, CLVs are the true, generally non-orthogonal, directions of growth and decay in the tangent space  \citep{ginelli_characterizing_2007,kuptsov_theory_2012,ginelli_covariant_2013}. Unlike the GS vectors, CLVs can uncover deep insights into flow instabilities. Most importantly, CLVs also provide a means to characterise hyperbolicity, a cornerstone concept in dynamical systems theory. Many theoretical results rely on the assumption of uniform hyperbolicity, for example  the  shadowing lemma \citep{pilyugin1999}. 

Several studies have applied Lyapunov analysis to flow systems, explored the structure of CLVs, and assessed the degree of hyperbolicity. \cite{CLVsofTurbulentChannelFlow_Nikitin_2018} characterized the leading CLV of a turbulent channel flow. \cite{qiqi_ferdandez_LEspectrum_airfoils} and \cite{ni_hyperbolicity_2019} computed the CLVs of the  2D flow around a NACA 0012 airfoil and the 3D flow around a circular cylinder, respectively. In both papers, the primary goal was to assess the hyperbolicity. \cite{xuandpaul2016} computed and visualised the CLVs of Rayleigh-Bénard convection, and examined the spatial power spectra. 

In the present study, we examine the stability of a chaotic flow around two square cylinders using Lyapunov analysis. We calculate the LEs and CLVs, analyze the unstable CLVs using Spectral Proper Orthogonal Decomposition (SPOD) and compare the obtained dominant structures with the eigenvectors obtained from GLSA of the time-average flow. To the best of our knowledge, this is the first work to apply flow decomposition methods such as SPOD to CLVs. Moreover, this is the first study to explicitly compare results from Lyapunov stability analysis with GLSA for chaotic flows. We also study the hyperbolicity of the system by computing the angles between pairs of CLVs.

The paper is structured as follows. The flow configuration and computational setup are described in section \ref{sec:flow_description_section}, followed by section \ref{sec:flow_characteristics} on the analysis of main flow characteristics. In section \ref{sec:LEs_and_CLVs}, we calculate the LEs and CLVs to characterize the chaotic dynamics and examine potential violations of hyperbolicity. We further characterise the unstable CLVs in section \ref{sec:SPOD_of_unstable_CLVs} using SPOD; this reveals the most dominant flow structures in the tangent space and their frequencies. Finally, in section \ref{sec:global_LSA} we perform GLSA of the time-averaged flow to identify the most unstable eigenmodes and compare these with the SPOD results of the unstable CLVs from the Lyapunov analysis. We summarize the main findings in section \ref{sec:conclusions} and also provide some thoughts on future research directions. 

\section{Flow configuration and computational details} \label{sec:flow_description_section}

\begin{figure}
\centerline{\includegraphics[width=1\linewidth]{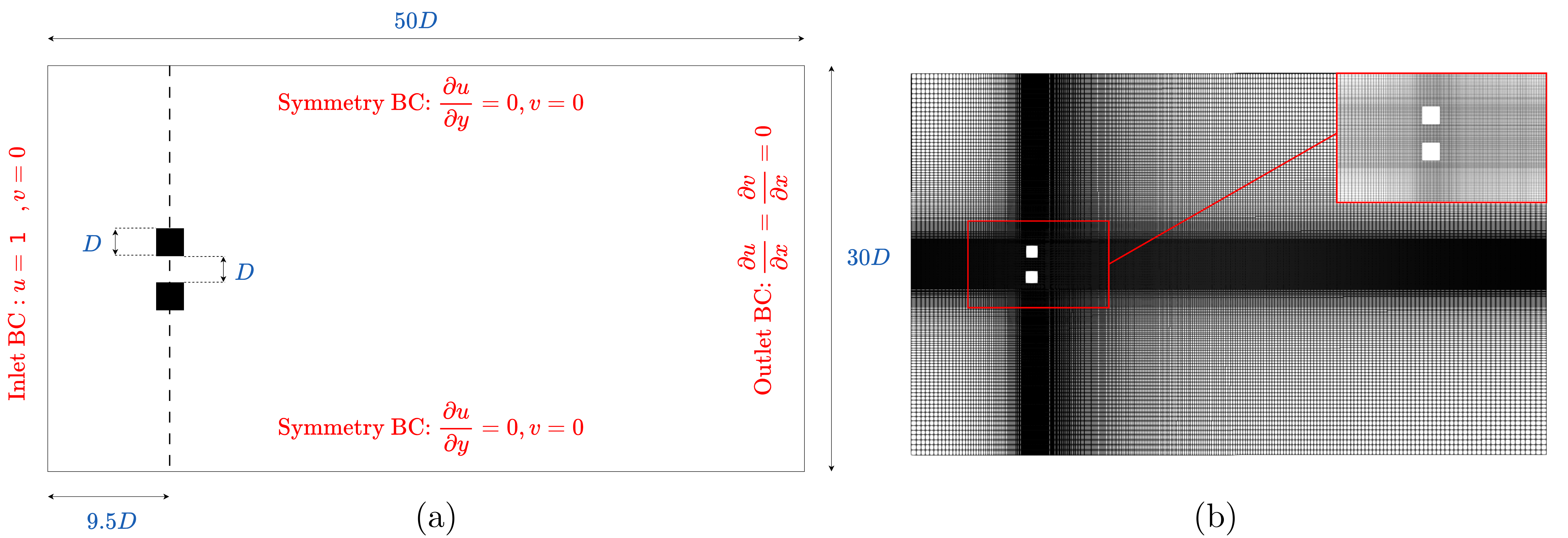}}
\caption{ (a) Flow configuration and boundary conditions, (b) Global and zoomed-in views of mesh 2.} 
\label{fig:gridgeometry}
\end{figure}

We consider the two-dimensional incompressible flow across two side-by-side square cylinders (prisms) separated by a gap distance equal to the prism side $D$, see figure \ref{fig:gridgeometry}(a). The non-dimensional form of the Navier-Stokes equations reads,
\begin{equation}
\begin{aligned}
\frac{\partial \boldsymbol{u}}{\partial t} + \boldsymbol{u} \cdot \nabla \boldsymbol{u} & = - \nabla p + \frac{1}{Re} \Delta \boldsymbol{u}, \\
\boldsymbol{\nabla} \cdot \boldsymbol{u} & = 0,
\end{aligned}
\label{kol_eq}
\end{equation}
\noindent where \( \boldsymbol{u} = (u,v) \) is the velocity vector with components $u$ and $v$ in the streamwise ($x$) and cross-stream directions ($y$) respectively,  $p$ is the pressure, and $\Delta$ the Laplacian operator. The reference quantities used for non-dimensionalization are \(D\) for the spatial variables, the free-stream velocity \(U_{\infty}\) for velocities, and \( \rho U_{\infty}^2 \) for pressure (where \( \rho \) is the fluid density). In the following, an overbar denotes a time-average quantity and a prime the fluctuation, for example $u=\overline{u}+u^\prime$ denotes the Reynolds decomposition of the instantaneous streamwise velocity $u$. The origin of the coordinate system $(x,y)$ is located at the centerline, in the gap between the cylinders, midway across the side length $D$.

The Reynolds number is defined as $\operatorname{Re} = U_{\infty} D / \nu$, where \( \nu \) is the kinematic viscosity of the fluid. We simulate the flow at $Re=200$. At this value, the vortex shedding from the top and bottom cylinders is temporally irregular, \citet{ma_wake_2017}. The non-dimensional frequency is defined as $St=f D/U_{\infty}$.

The equations are discretised using the finite-volume methodology applied to a Cartesian mesh. Uniform velocity is imposed at the inlet, zero pressure gradient at the outlet, and symmetry conditions at the top and bottom boundaries. The Crank-Nicolson scheme is employed for time marching with time step $\delta t=0.01$. The convective and viscous terms are discretised using a second-order central scheme. The flow was simulated for $500$ time units (one time unit is equal to $D/U_{\infty}$) until a fully developed flow was obtained. The simulation was then restarted and continued over \( 10000 \) time units, during which data were collected for further processing.

\begin{table}
\centering
\renewcommand{\arraystretch}{1.3} 
\begin{tabular}{lcccccccc}
\hline
\text{} & \text{Cell count} & \text{Cell thickness } & \text{Cells across} & $\overline{C_D}$ & $\overline{C_L}$ & $\left(C_D\right)_{\text{rms}}$ & $\left(C_L\right)_{\text{rms}}$ \\
\text{} & \text{} & \text{next to the wall} & \text{prism side} & \text{} & \text{} & \text{} & \text{} \\
\hline
Mesh 1 & 66{,}258 & 0.02D & 20 & 2.155 & 0.209 & 0.351 & 0.775 \\
Mesh 2 & 84{,}253 & 0.01D & 30 & 2.049 & 0.204 & 0.338 & 0.762 \\
\citet{ma_wake_2017} & - & 0.01D & 25 & 2.03 & 0.19 & - & - \\
\hline
\end{tabular}
\caption{Mesh details and force coefficients for the bottom cylinder.}
\label{table:mesh_convergence}
\end{table}
 
To assess the accuracy of the simulation, a mesh independence study was conducted. Two meshes, 1 and 2, consisting of 66,258 and 84,253 cells respectively, were evaluated. The finer mesh, shown in figure \ref{fig:gridgeometry}(b), has higher cell density around the prisms and the near field. Table \ref{table:mesh_convergence} summarizes relevant mesh details and the computed mean and rms values of the lift ($C_L$) and drag ($C_D$) coefficients for the bottom cylinder. Comparison with the reference values of  \citet{ma_wake_2017} confirms the accuracy of the simulation, especially with the finer mesh. 

\begin{figure}[ht!]
\centerline{\includegraphics[width=1\linewidth]{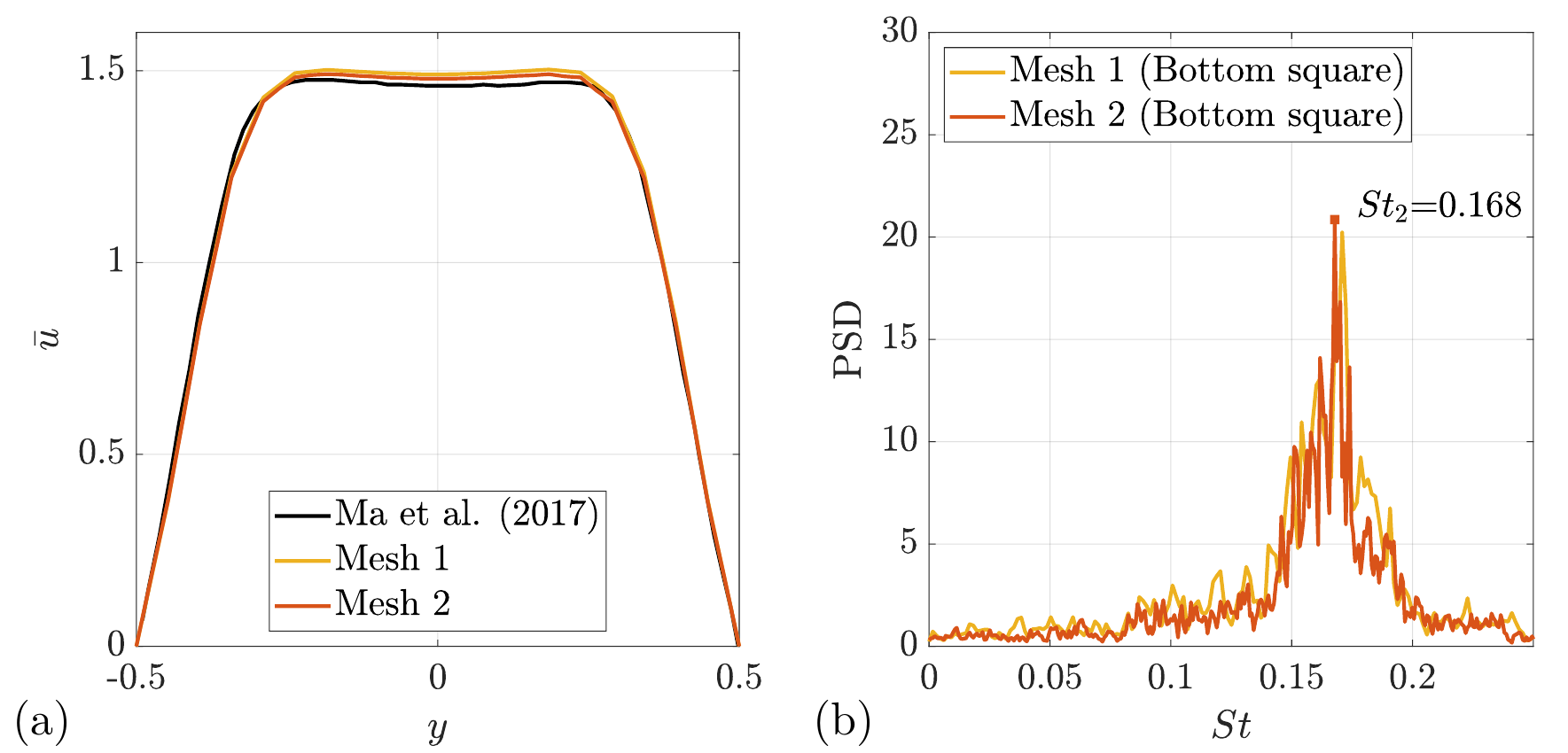}}
  \caption{Mesh independence study (a) $\bar{u}(y)$ across the gap between the cylinders at $x=0$, (b) Spectrum of $C_L'$ for the bottom cylinder.}
\label{fig:mesh_convergence}
\end{figure}

Figure \ref{fig:mesh_convergence}(a) shows comparison of the $\bar{u}(y)$ velocity profile across the gap at $x = 0$ with that of \citet{ma_wake_2017}. Panel (b) shows the frequency spectrum of the lift coefficient (\(C_L\)) for the bottom cylinder. There is a clear peak at frequency $St=0.168$; this is further analyzed below. The velocity profiles and the spectra exhibit minimal differences between the two meshes, suggesting that the solution is indeed grid independent. In the following, the flow fields from mesh 2 are used.

\section{Flow characteristics \label{sec:flow_characteristics}}

\begin{figure}[ht!]
\centerline{\includegraphics[width=1.1\linewidth]{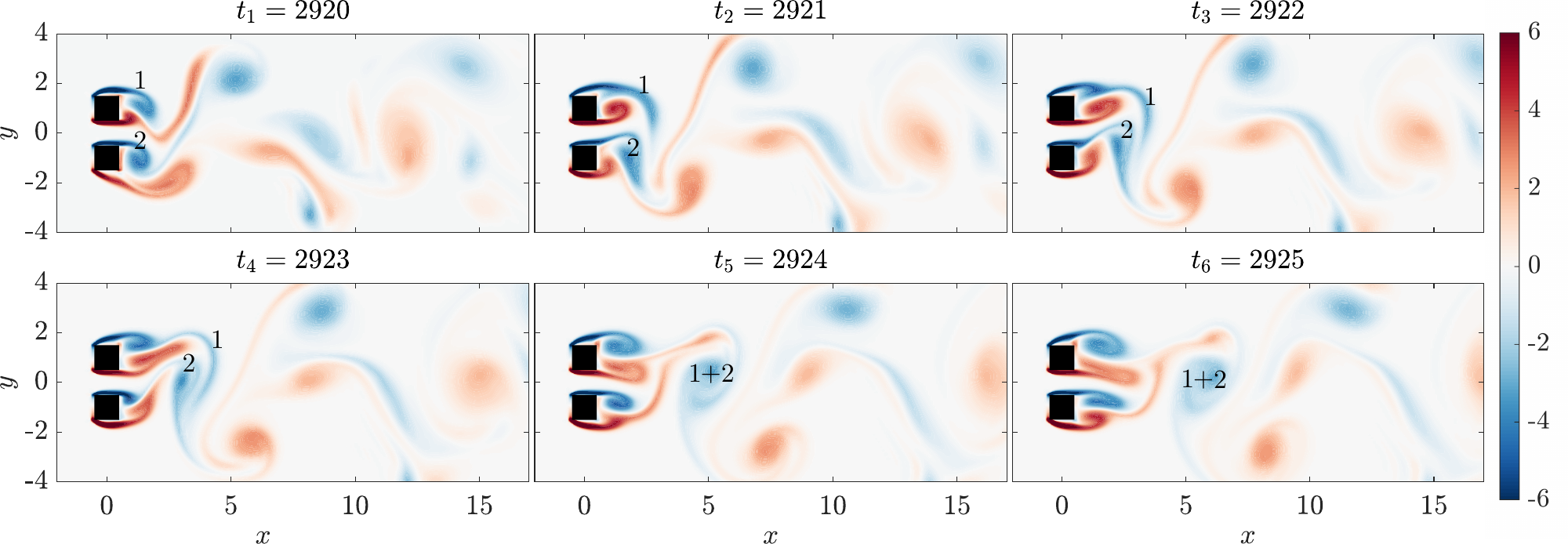}}
  \caption{Contour plots of instantaneous vorticity ($\nabla \times \boldsymbol{u}$) at 6 time instances. The plots depict the flapping motion of the jet and the merging of vortices 1 and 2. For an animation of vorticity contours, see supplementary video.} 
\label{fig:flappingjet}
\end{figure}

Figure \ref{fig:flappingjet} depicts instantaneous vorticity snapshots at six time instants $t_1 \dots t_6$. A jet emanates from the gap between the cylinders and disrupts the periodic vortex shedding behind each cylinder, leading to irregular flow patterns (the flow is in fact chaotic, as will be demonstrated in section \ref{sec:LEs_and_CLVs} below). The jet exhibits a flapping motion which is determined by the pressure field induced by the vortices shed from the prisms faces on either side of the gap. In the vorticity snapshots of the top row the jet bends mainly upwards, while in bottom row it bends downwards. Note how the vortices are stretched by the flow, from highly concentrated blobs to long vorticity filaments. Notice also the long excursions of the wake in the cross-stream direction. Vortex filaments can reach $y$ values larger than $4$, this is due to the sweeping motion of the flapping jet.

Close examination of the vorticity sequences reveals an interesting phenomenon. To visualize it, two clockwise-rotating vortices, marked with numbers 1 and 2, are tracked. The vortices are shed from the top face of each prism. At time instant $t_1$ the two vortices are highly concentrated and almost in phase. At the subsequent time instants $t_2$ and $t_3$, vortex 1 is stretched and bends downwards coming in close proximity with vortex 2 which at the same time moves upwards. At $t_4$, the two vortices have moved downstream together and start to merge. The merging process is finished at $t_5$, and a single vortex is formed that is carried downstream, see instant $t_6$. This process of vortex merging is an important characteristic of this flow. It can be detected in the spectra and the Lyapunov covariant vectors, as will be seen later. 

\begin{figure}[ht!]
    \centering
    \includegraphics[width=1\linewidth]{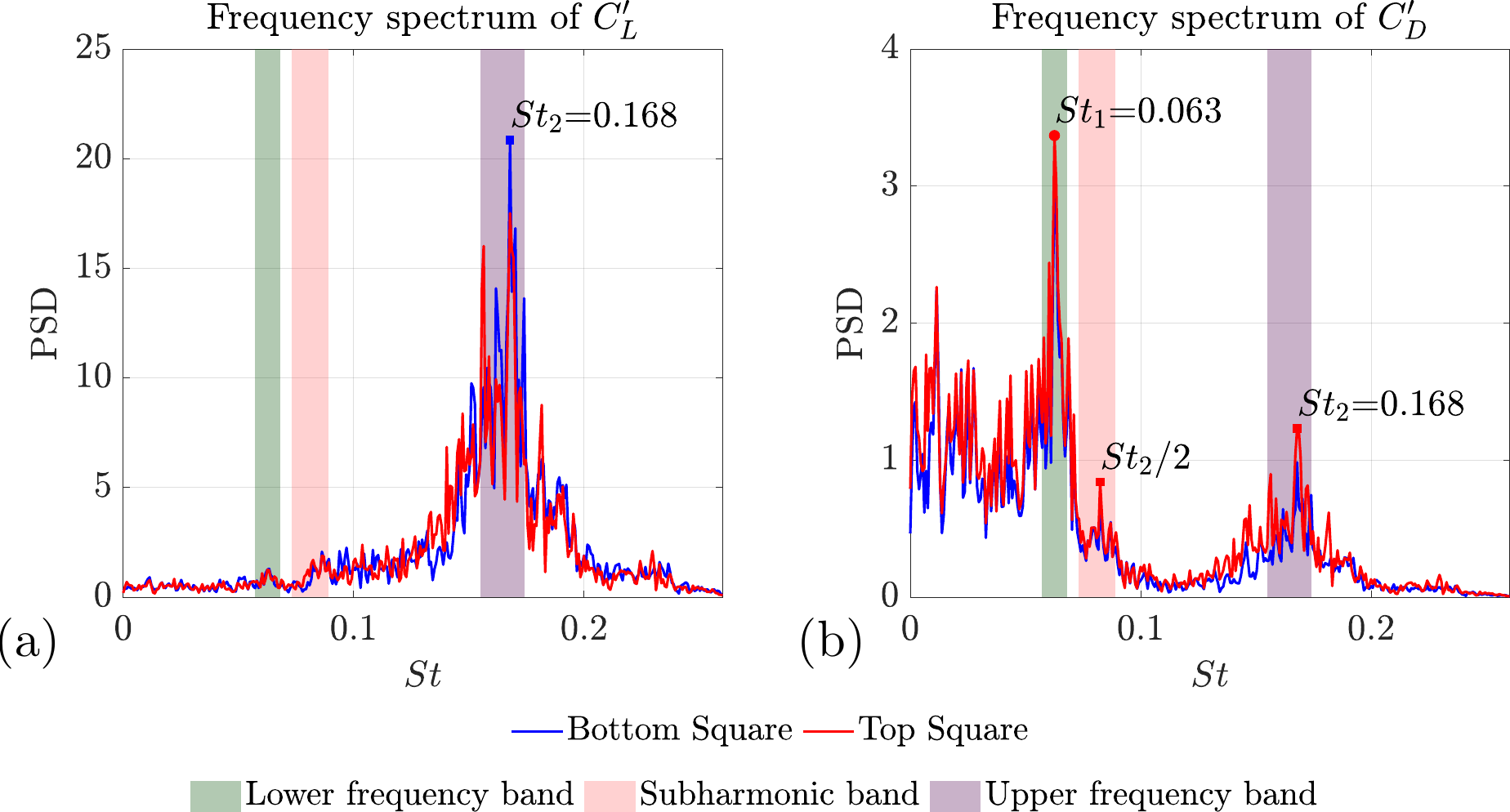}
    \caption{Frequency spectrum of the fluctuating force coefficients for (a) $C_L^\prime$ and (b) $C_D^\prime$ of the two prisms. 
    }
    \label{fig:CDCL_Spectrum_bothsquares}
\end{figure}

The flow irregularity leaves its footprint on the spectra of the fluctuating \(C_L^\prime\) and \(C_D^\prime\) coefficients shown in figure \ref{fig:CDCL_Spectrum_bothsquares}. The spectra were computed with Welch’s method using eight segments with 20\% overlap and a Hamming window on each segment. They are almost identical for the top and bottom prisms, as expected due to symmetry. This also indicates that the signal is long enough that can capture the matching of the spectra at low frequencies. 
In the \(C_D\) spectrum there are two peaks at \(St_{1}=0.063\) and \(St_{2}=0.168\) (the former stronger than the latter), whereas the \(C_L\) spectrum exhibits only one peak at \(St_{2}=0.168\).  Note however that multiple smaller peaks appear, especially around \(St_2\), due to the flow irregularity. Two main frequency bands can be identified, a lower band (\(St\in[0.057,0.068]\)) around \(St_{1}\) and an upper band (\(St\in[0.155,0.174]\)) around \(St_{2}\). Notice also the presence of a weaker peak at the subharmonic \(St_2/2=0.084\) in the \(C_D\) spectrum; a third band \(\bigl(St\in[0.073,0.089]\bigr)\) is marked around it. The bandwidths are determined by the full width at half maximum (FWHM) of the corresponding spectral peak, see \cite{Bhattacharya}.

\begin{figure}[ht!]
    \centering
    \includegraphics[width=1\linewidth]{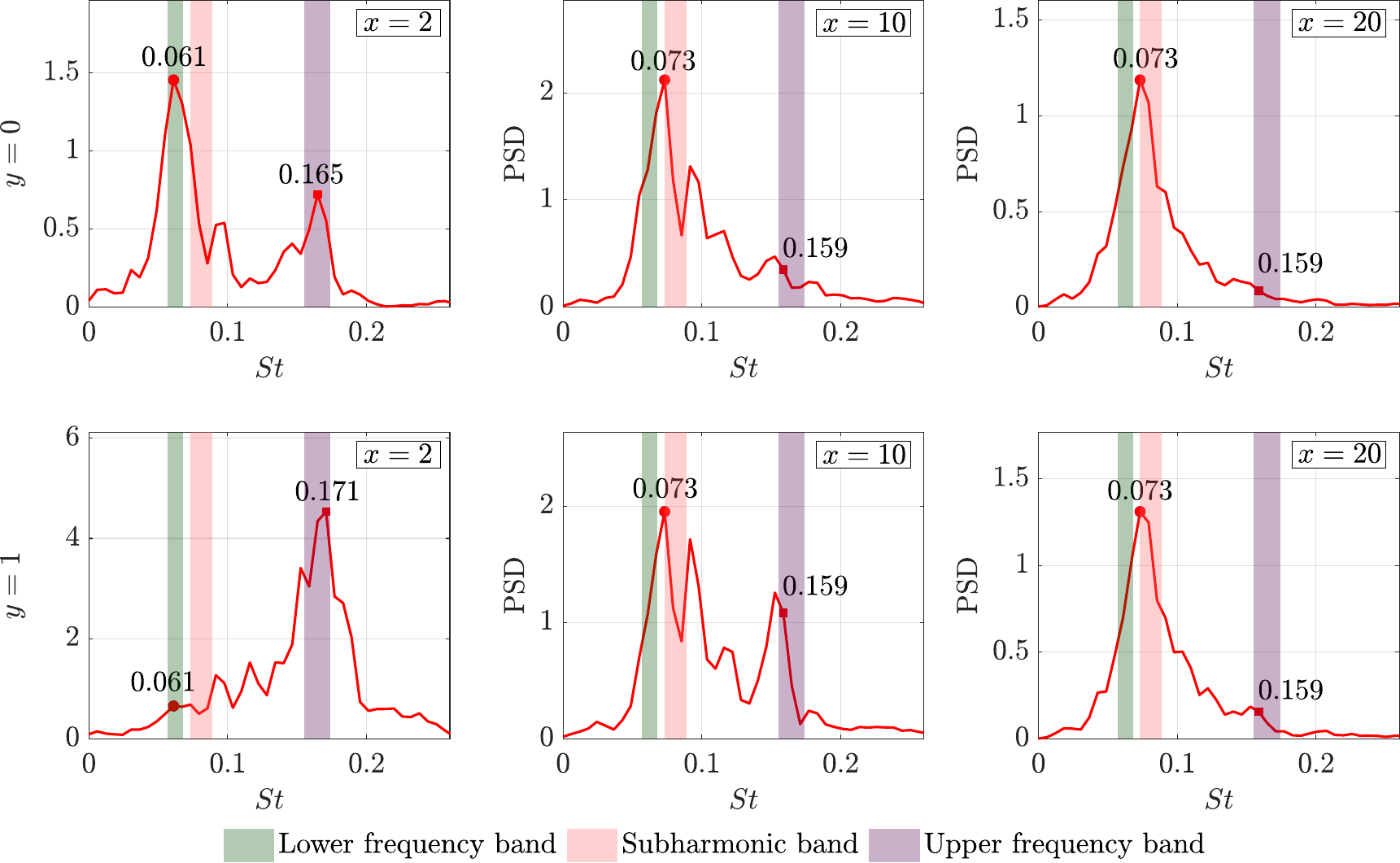}
\caption{Spectra of the transverse velocity $v(x,y)$ at $x=2$ (left column), $10$ (middle column) and $20$ (right column). Top row: along the gap centerline ($y=0$); bottom row: along the centerline behind the top prism ($y=1$). The shaded regions mark the frequency bands defined in figure \ref{fig:CDCL_Spectrum_bothsquares}.}
    \label{fig:pointwise_spectra}
\end{figure}

To explain the physical mechanisms behind $St_1$ and $St_2$, we compute the spectra of the transverse velocity $v(x,y)$ at three locations $x=2$, $10$ and $20$ along $y=0$ (gap centerline) and $y=1$ (centerline behind top square cylinder). The results are shown in figure \ref{fig:pointwise_spectra}. At point $x=2,\ y=0$ the flow physics is determined by the jet  motion and the PSD exhibits a pronounced peak at $0.061$ which is very close to $St_1$, indicating that the latter corresponds to the low-frequency flapping motion. It is interesting to note that the prominent peak in the $C_D$ spectrum is therefore caused by the gap-jet flapping. At the same streamwise location but at $y=1$, the PSD peaks at $0.171$, which is close to $St_2$, indicating that the latter represents the primary vortex shedding frequency in the wake behind the individual prisms. Thus vortex shedding explains the main peak in $C_L$ spectrum and the secondary peak in the $C_D$ spectrum. It is very interesting to note that further downstream ($x=10$ and $20$) at both $y$ lines, the PSD peaks converge to a single peak at $0.073$, which is within the subharmonic frequency band. 

\begin{figure}[ht!]
    \centering
    \includegraphics[width=1\linewidth]{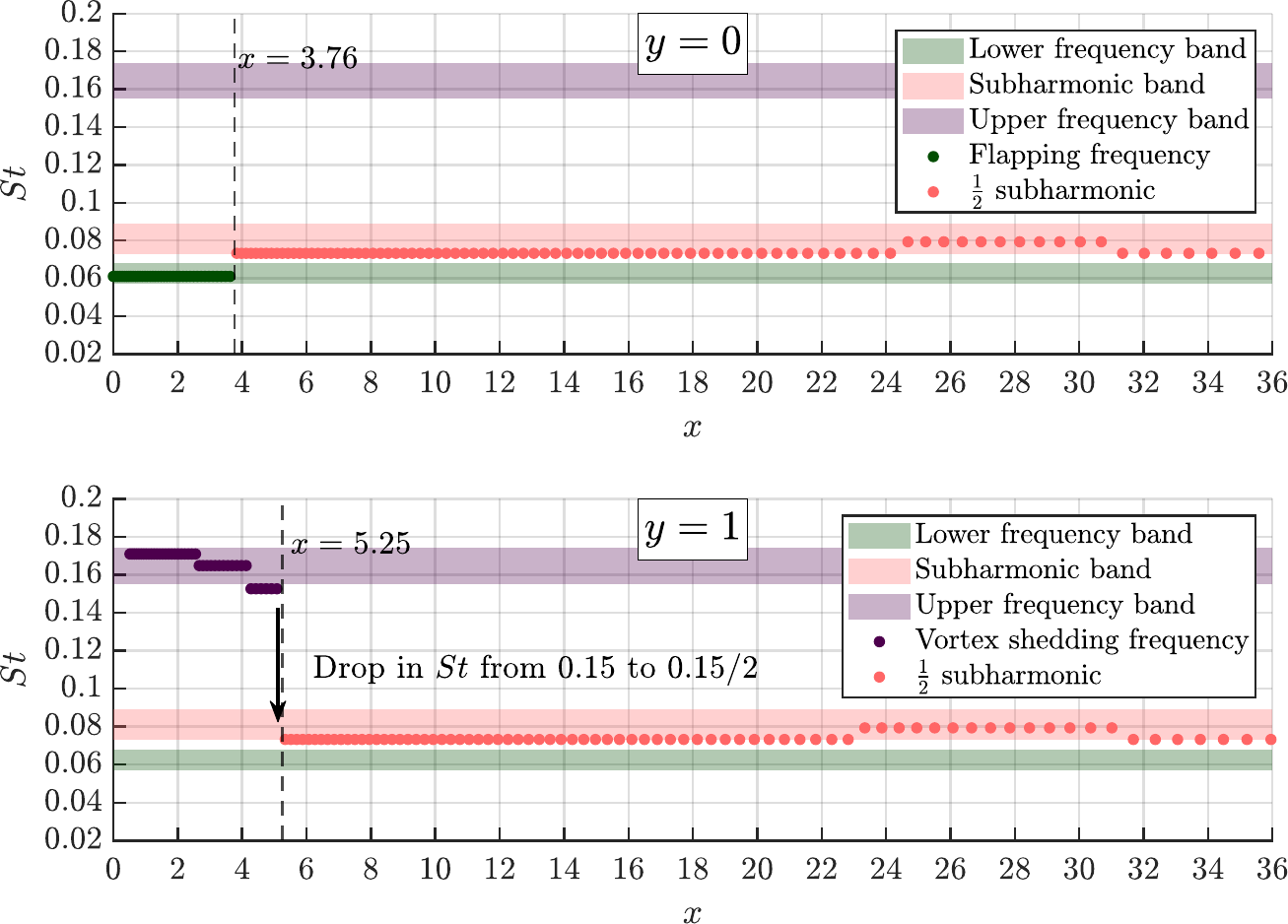}
    \caption{Streamwise evolution of the dominant $St$ of the $v$ velocity spectrum along (a) $y=0$ and (b) $y=1$. The shaded regions mark the frequency bands defined in figure \ref{fig:CDCL_Spectrum_bothsquares}.}
    \label{fig:subharmonic_plot}
\end{figure}

Figure \ref{fig:subharmonic_plot} shows the dominant $St$ of the PSD of $v(x,y)$ along $x$ for both $y$ lines. In the near wake ($x\lesssim4$), the dominant frequency is within the upper frequency band (i.e.\ close to $St_{2}$) at $y=1$ and within the lower band (i.e.\ close to $St_{1}$) at $y=0$, confirming the different mechanisms behind each band. However, downstream of $x \approx 5$ the dominant $St$ drops to approximately $St_{2}/2$, consistent with the subharmonic generated by vortex pairing events such as the one shown in figure \ref{fig:flappingjet}. The spatial location of the frequency drop matches very closely with the merging location depicted in the aforementioned figure. This frequency shift marks the onset of wake dynamics dominated by paired vortices and the decay of the gap‐jet oscillation. The association of vortex merging with the subharmonic has been reported repeatedly in the literature, see \cite{Meiburg_1987, Shaabani-Ardali_Sipp_Lesshafft_2019}. 

\begin{figure}[ht!]
    \centering
    \includegraphics[width=0.8\linewidth]{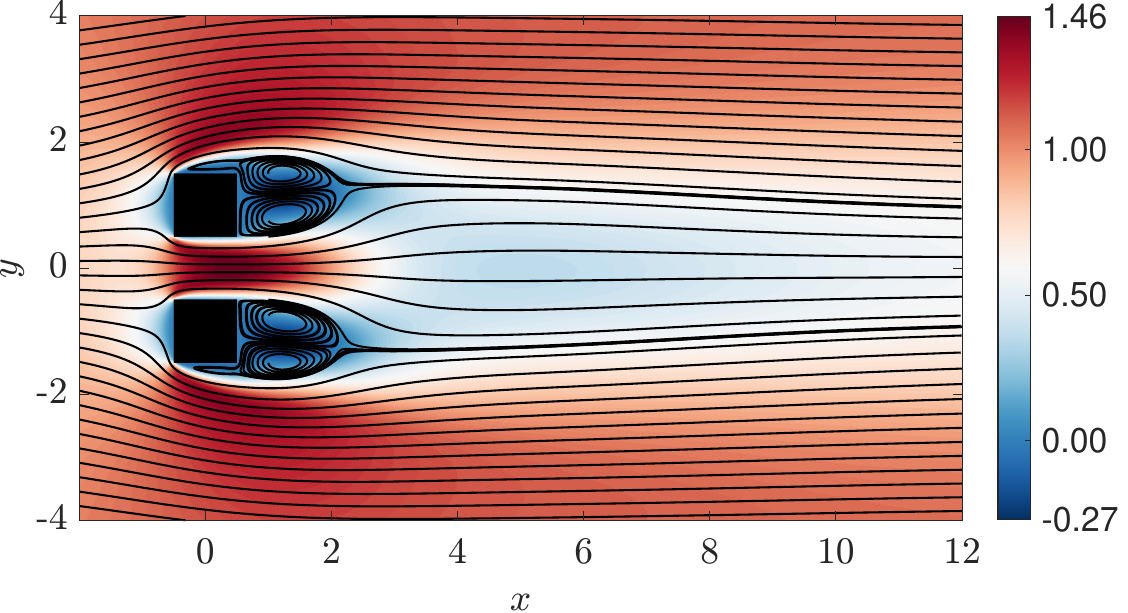}
    \caption{Contours of the time-averaged streamwise velocity $\bar{u}$ superimposed to streamlines.}
    \label{fig:timeaveraged_baseflowU}
\end{figure}

Despite the underlying irregular dynamics, the time-averaged streamwise velocity field $\bar{u}$, shown in figure \ref{fig:timeaveraged_baseflowU}, is symmetric about $y=0$. A high-speed jet emerges from the gap and two symmetric recirculation zones form immediately downstream of the cylinders. The coexistence of the jet and the recirculation zones results in a highly non-uniform velocity field in the very near wake. Further downstream the flow behind the cylinders recovers and the jet velocity decays, making the velocity profiles in the cross-stream direction more uniform.

\subsection{Spectral Proper Orthogonal Decomposition (SPOD) of the flow}
To obtain more insight, SPOD \citep{towne_spectralPOD_2018} is applied to the fluctuating flow field. The method will identify the structures beating at the dominant frequencies identified earlier. A total of \(K = 100,000\) snapshots of velocity perturbations obtained every $\delta t=0.01$ are stacked column by column to form matrix \({Y}\),
\begin{equation}
Y = \left[\begin{array}{cccc}
{u^{\prime}_1}^{(1)} & {u^{\prime}_1}^{(2)} & \cdots & {u^{\prime}_1}^{(K)} \\
{v^{\prime}_1}^{(1)} & {v^{\prime}_1}^{(2)} & \cdots & {v^{\prime}_1}^{(K)} \\
\vdots & \vdots & \cdots & \vdots \\
{u^{\prime}_N}^{(1)} & {u^{\prime}_N}^{(2)} & \cdots & {u^{\prime}_N}^{(K)} \\
{v^{\prime}_N}^{(1)} & {v^{\prime}_N}^{(2)} & \cdots & {v^{\prime}_N}^{(K)}
\end{array}\right].
\end{equation}
\noindent The superscripts denote the snapshot index \(k = 1, 2, \dots, K\) and the subscripts the cell number \(n = 1, 2, \dots, N\), where $N=84,253$ (cell count of mesh 2). Matrix $Y$ is divided into 11 blocks with a 50\% overlap. For each frequency $St_k$, a cross-spectral density (CSD) matrix \( \mathbf{S}(St_k) \) is formed and the SPOD eigenvalue problem is solved to find the SPOD modes \( \Phi^{st_k} \) and their energies \( \lambda(St_k) \).  Since the CSD matrix is Hermitian, the eigenvalues are real but the eigenvectors (SPOD modes) are complex, i.e.\ $ \Phi^{St_k}  = \Phi_r^{St_k} + i \Phi_i^{St_k}$; for more details see \cite{towne_spectralPOD_2018}.

\begin{figure}
    \centering
    \includegraphics[width=0.75\linewidth]{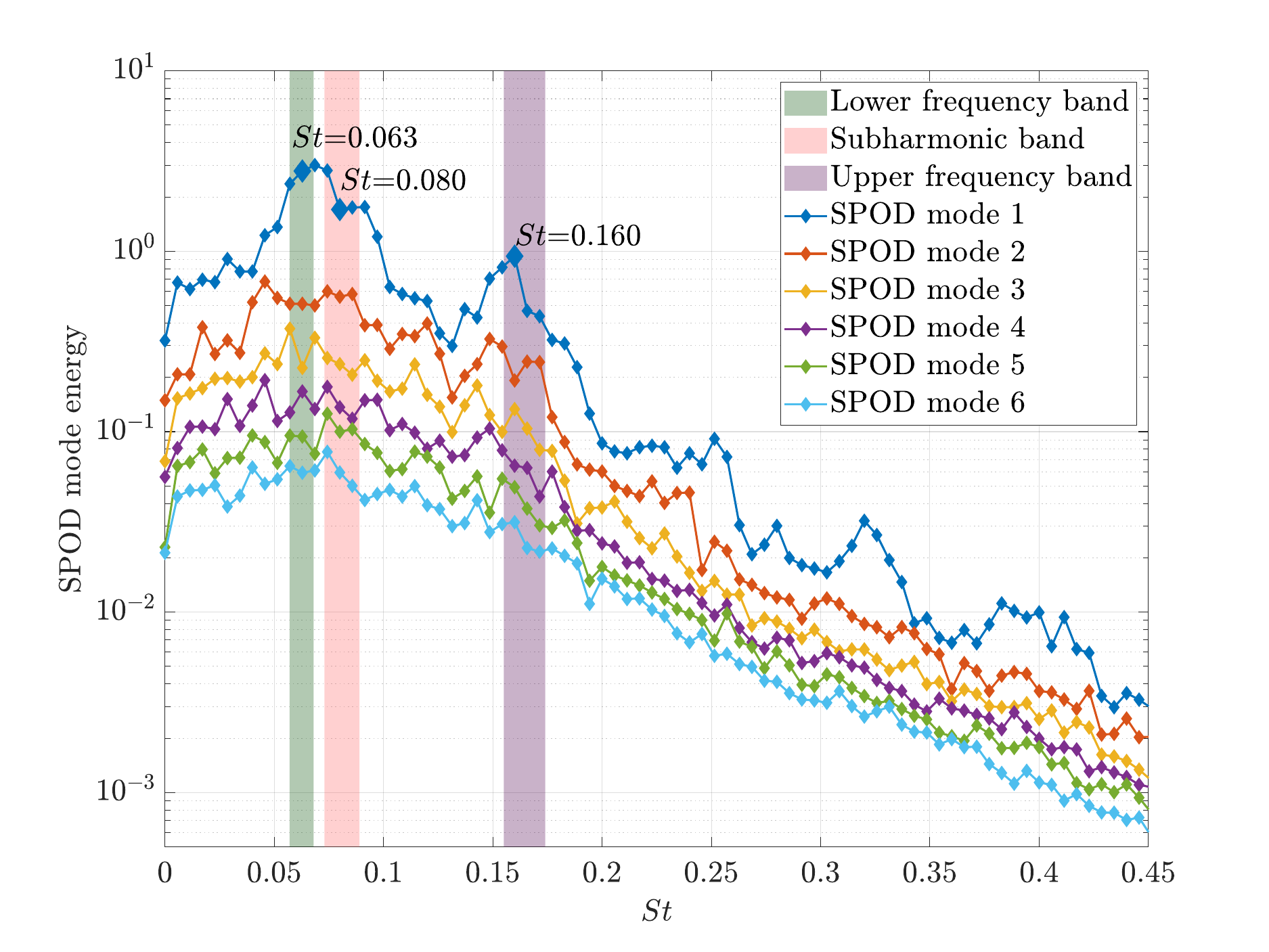}
    \caption{Energy spectrum of the first 6 SPOD modes of the flow.}
    \label{fig:SPOD_nonlinear_ES}
\end{figure}

Figure~\ref{fig:SPOD_nonlinear_ES} presents the energy spectrum of the first six SPOD modes. Mode 1 exhibits two clear peaks at $0.063$ and $0.160$, that match closely $St_1$ and $St_2$ respectively of the $C_D$ spectrum. For the lower frequency band, the energy of mode 1 is markedly higher than that of mode 2, indicating that the gap‐jet flapping motion is well approximated by just one mode.  In the higher frequency band, the energy separation between modes 1 and 2 is not so pronounced. Note also the presence of a band with $0.080$, the subharmonic of the $0.160$. 

\begin{figure}
    \centering
    \includegraphics[width=1\linewidth]{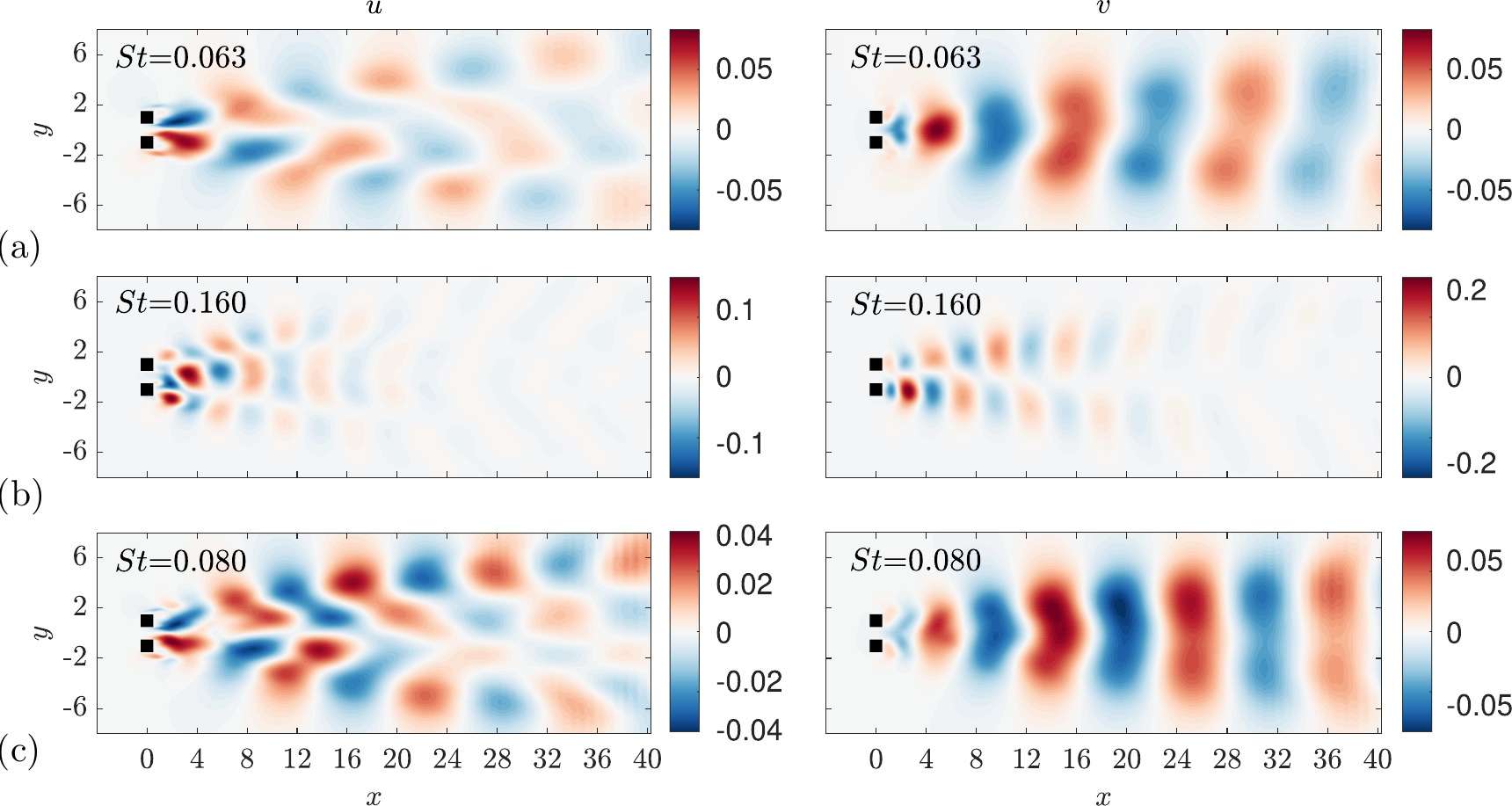}
    \caption{Contour plots of streamwise (left column) and cross-stream (right column) velocities of the real part of SPOD mode 1  at \( St = 0.063 \), $\Phi_{1r}^{0.063}$ (a), at \( St = 0.160 \),  $\Phi_{1r}^{0.160}$ (b) and at \( St = 0.080 \), $\Phi_{1r}^{0.080}$ (c).}
    \label{fig:SPOD_nonlinear}
\end{figure}

Figure~\ref{fig:SPOD_nonlinear} shows contours of the streamwise ($u$) and transverse ($v$) velocities of the real part of SPOD mode 1 at $St=0.063$, $0.160$, and the subharmonic $0.080$. At $St = 0.063$ the $v$ contours peak at the gap centerline in the near wake. Both $u$ and $v$ reach their maxima at $x \approx 4$, precisely where the jet-flapping motion is most discernible. At $St = 0.160$ (the vortex‐shedding frequency), a train of small-scale structures appears in the wake of each prism in the $v$ contours. The peaks now appear behind the prisms, not in the gap centerline. Notice that the peaks are displaced further away from the centerline as $x$ increases, indicating the wake expansion. The $v$ field is antisymmetric with respect to the gap centerline but the $u$ field is symmetric. The $v$ mode attain maximum values in the near wake region ($x<4$), as for $St=0.063$. On the other hand, the $v$ contours of the subharmonic at $St=0.080=0.160/2$) grow rapidly in the near wake (where vortex merging occurs), peaks in the mid‐wake $x \in [12,20]$ and then decays. 

Overall, the dominant SPOD mode captures the three key phenomena identified earlier: jet-flapping, vortex shedding behind each cylinder, and vortex pairing. As can be seen in the SPOD energy spectrum, gap‐jet-flapping carries the largest energy, followed by vortex pairing and then asynchronous shedding.

Having characterized the main features of the flow, the next step is to consider the flow as a dynamical system and analyse the Lyapunov exponents (LEs) and covariant Lyapunov vectors (CLVs). A key objective is to explore the footprint of the above features onto the CLVs.

\section{Analysis of the flow from a dynamical systems perspective}\label{sec:LEs_and_CLVs}

\subsection{Lyapunov exponents}
After discretisation, the system of governing equations \eqref{kol_eq} can be put in the general form, 
\begin{equation}
\frac{d \mathbf{u}}{dt} = \mathbf{f}\left(\mathbf{u}\right), \quad \mathbf{u}(0) = \mathbf{u}_0
\label{eq:L1}
\end{equation}
\noindent where \(\mathbf{u}(t)\) is the state vector $\mathbf{u}(t) = \{u_1(t), v_1(t), \ldots, u_{N}(t), v_{N}(t)\} \in \mathbb{R}^{2N}$, and $N$ is the number of cells. To get \eqref{eq:L1}, we have used the continuity equation to eliminate pressure. This is done by  constructing a Poisson equation for pressure \citep{pope_2000} which can be notionally solved, and thus $\nabla P$ can be expressed in terms of velocities. As the dynamical system evolves, the state vector traces a trajectory in phase space. We define the evolution operator $\mathcal{L}^{t}$ that maps $\mathbf{u}(0)$ to $\mathbf{u}(t)$ under the dynamics \eqref{eq:L1}, $\mathcal{L}^{t}(\mathbf{u}(0)) =\mathbf{u}(t)$.

A key concept in dynamical systems is ergodicity, which implies that the time-average state of the system is independent of the initial condition \citep{Birkhoff}. For ergodic systems, the long-time trajectory converges to a bounded region in phase space known as attractor \citep{Ruelle1980}. Attractors can be classified into three types: fixed (or equilibrium) points, limit cycles, and strange attractors. Fixed-point attractors are stable points that trajectories converge to over time, limit-cycle attractors correspond to stable periodic orbits, and strange attractors represent chaotic systems that exhibit complex, irregular patterns (as the one considered in this study).

We define our (assumed ergodic) dynamical system in the \(2N\)-dimensional Riemannian manifold \(\mathcal{M}\). The \cite{Oseledets1968} theorem establishes that a set of Lypunov exponents (LEs) exists, provided there is a tangent space \(\mathcal{T}_\mathbf{u} \mathcal{M}\)  that describes the local geometry of the attractor for every state $\mathbf{u}(t)$. $\mathcal{T}_\mathbf{u} \mathcal{M}$ is a $2N$-dimensional vector space consisting of all the possible directions in which the state $\mathbf{u}(t)$ can be perturbed. To study the response of a dynamical system to an arbitrary perturbation \(\bm{\delta \mathbf{u}} \in \mathcal{T}_\mathbf{u} \mathcal{M}\), we define the local expansion rate as, 
\begin{equation}
\gamma \left( \bm{\delta \mathbf{u}}, \mathbf{u},t \right) = \frac{\left\|D \mathcal{L}^t(\mathbf{u}) \bm{\delta \mathbf{u}} \right\|}{\| \bm{\delta \mathbf{u}}\|}
\end{equation}
\noindent where \(\|\cdot\|\) denotes the Euclidean norm and $D \mathcal{L}^t(\mathbf{u})$ is the linear evolution operator of the tangent space. More specifically,  $D \mathcal{L}^t(\mathbf{u})$ evolves the linearised form of \eqref{eq:L1} around $\mathbf{u}$,   
\begin{equation}
\frac{\partial (\bm{\delta \mathbf{u}})}{\partial t}= \frac{\partial  \mathbf{f} }{\partial  \mathbf{u} }  \delta \mathbf{u},  
\label{kol_lin1_eq}
\end{equation}
over a time window $t$. In other words, if the perturbation at the start of the time window is $\bm{\delta \mathbf{u}}$, the notation $D \mathcal{L}^t(\mathbf{u}) \bm{\delta \mathbf{u}}$ is the result of the action of this operator, i.e.\  $D \mathcal{L}^t(\mathbf{u}) \bm{\delta \mathbf{u}}=\bm{\delta \mathbf{u}}(t)$. Thus the operator  performs the mapping $\mathcal{T}_\mathbf{u} \mathcal{M} \rightarrow \mathcal{T}_{\mathcal{L}^t(\mathbf{u})} \mathcal{M}$. For the Navier-Stokes equations, the linearized form is,  
\begin{equation}
\begin{aligned}
\frac{\partial (\bm{\delta \mathbf{u}})}{\partial t}+ \mathbf{u} \cdot \nabla (\bm{\delta \mathbf{u}}) +\bm{\delta \mathbf{u}} \cdot \nabla \mathbf{u} & =-\nabla (\delta p)+\frac{1}{Re} \Delta (\bm{\delta \mathbf{u}}), \\
\nabla \cdot (\bm{\delta \mathbf{u}}) & =0,
\end{aligned}
\label{kol_lin_eq}
\end{equation}
\noindent where $\bm{\delta \mathbf{u}}= (\delta u, \delta v)$.

The Lyapunov exponent \(\lambda\) associated with the perturbation vector \(\bm{\delta \mathbf{u}}\) is the long-time average,
\begin{equation}
\lambda (\bm{\delta \mathbf{u}})  = \lim_{t \rightarrow \infty} \frac{1}{t} \ln \frac{\left\|D \mathcal{L}^t(\mathbf{u}) \bm{\delta \mathbf{u}}\right\|}{\| \bm{\delta \mathbf{u}}\|}
\label{lambda_expression}
\end{equation}
This exponent represents the average exponential rate of expansion (or contraction) of the perturbation \(\bm{\delta \mathbf{u}}\), i.e.\ \(\left\| \bm{\delta \mathbf{u}}(t)\right\| \sim \| \bm{\delta \mathbf{u}}(0)\| \mathrm{e}^{\lambda t}\). The \cite{Oseledets1968} theorem establishes that \(\lambda\) assumes a finite number of distinct values \(\lambda_1 > \lambda_2 > \cdots > \lambda_m\), where \(m \leqslant 2N\), as the perturbation vector \(\bm{\delta \mathbf{u}}\) varies in the tangent space \(\mathcal{T}_\mathbf{u} \mathcal{M}\).

The set of LEs, known as the 'Lyapunov spectrum', is a characteristic of the dynamical system. In a chaotic dynamical system there is at least one positive LE, i.e.\ \(\lambda_1 > 0\). This means that  the perturbed trajectory \( \bm{\mathbf{u}}(t)+\bm{\delta \mathbf{u}}(t)\) will diverge exponentially from the reference trajectory \( \bm{\mathbf{u}}(t)\) at an average rate $\lambda_1$.

When a randomly chosen set of \(m \leq 2N\) infinitesimal perturbations \((\bm{\delta \mathbf{u}}_1, \bm{\delta \mathbf{u}}_2, \ldots, \bm{\delta \mathbf{u}}_m)\) are propagated in the tangent space using \eqref{kol_lin1_eq} (or \eqref{kol_lin_eq} in our case), the angles between \(\bm{\delta \mathbf{u}}_i\) reduce with time as all perturbations align along the most unstable direction. Thus, one can compute only the leading LE, and identifying smaller LEs becomes numerically ill-conditioned. \cite{benettin_lyapunov_1980,benettin_lyapunov_1980-1} devised an algorithm in which the set of \(m\) perturbations is propagated for a time interval \(t_{\text{step}}\), the evolved ${\bm{\delta \mathbf{u}}}_i$ are then orthonormalised and replaced by a set of orthonormal vectors that span the same subspace. These vectors are used as initial conditions and are propagated for another time segment \(t_{\text{step}}\). \citet{benettin_lyapunov_1980} proved that using Gram-Schmidt QR decomposition a set of coordinate-independent LEs can be obtained. Coordinate independence means that the computed LEs are intrinsic to the system dynamics and therefore independent of the chosen coordinate system, and therefore the norm used to compute them, see also \cite{Bewley}. The process is presented in algorithm \ref{Benettin_algo} below.

\begin{algorithm}
\caption{Calculation of Lyapunov exponents \citep{benettin_lyapunov_1980}} \label{alg:lyapunov}
\begin{algorithmic}[1]
    \State Given a generic initial condition \(\mathbf{u}_0\), integrate  $\dfrac{d \mathbf{u}}{dt} = \mathbf{f}(\mathbf{u})$  until the trajectory has reached the ergodic attractor.
    \State Initialize orthonormal perturbations \(\{\bm{\delta \mathbf{u}}_j\}_{j=1 \dots m}\) and reset time to \(t = 0\).
    \State Divide the total time interval $[0,T]$ into \(K\) time segments, each of length \(t_{\text{step}}\).
    \For{\(i = 1\) to \(K\)}
        \State Evolve the phase space dynamics for 
            \(t_{\text{step}}\),   
        \( \mathbf{u}(t+t_{\text{step}})= \mathcal{L}^{t_{\text{step}}}(\mathbf{u}(t))\).
        \For{\(j = 1\) to \(m\)}
            \State Evolve the tangent space vector \(\bm{\delta \mathbf{u}}_j\) for 
            \(t_{\text{step}}\),  $\tilde{\bm{\delta \mathbf{u}}}_j=D \mathcal{L}^{t_{\text{step}}}(\mathbf{u}) \bm{\delta \mathbf{u}}_j$
        \EndFor
        \State Orthonormalize the subspace spanned by \([\tilde{\bm{\delta \mathbf{u}}}_1, \ldots, \tilde{\bm{\delta \mathbf{u}}}_m]\) with QR decomposition
        \State Store \(Q\) and \(R\) matrices (if  CLVs are to be also computed)
        \State Update \(\lambda_j\):
        \[
        \lambda_j  \to \lambda_j + \frac{1}{i \cdot t_{\text{step}}} \log \left(R_{jj}\right).
        \]
        \State Reinitialize the perturbations setting \(\{\bm{\delta \mathbf{u}}_j\}_{j=1 \dots m} = Q\).
        \State Set $t \to t+t_{\text{step}}$
    \EndFor
\end{algorithmic}
\label{Benettin_algo}
\end{algorithm}

\begin{figure}[ht!]
\centerline{\includegraphics[width=1\linewidth]{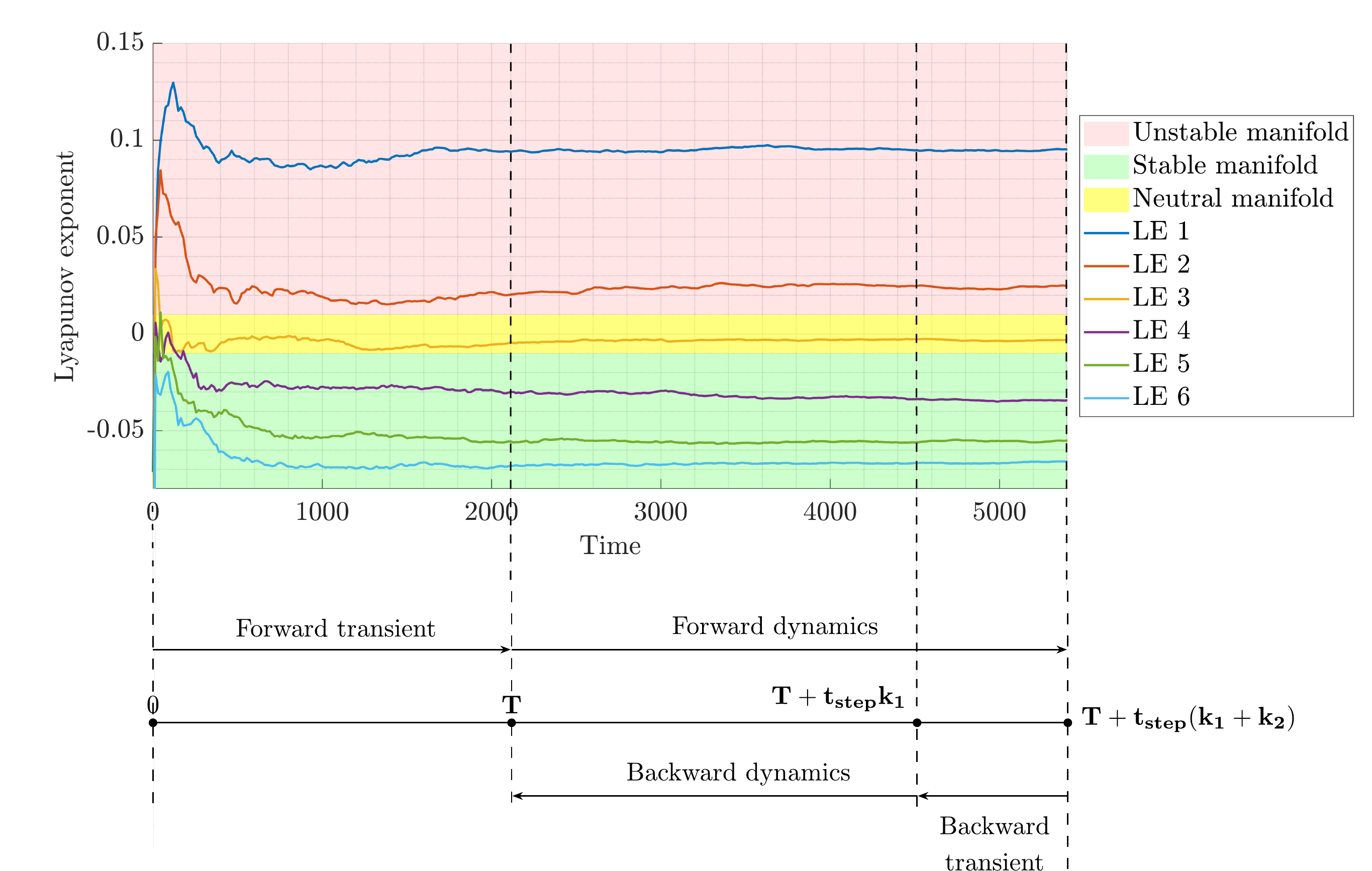}}
  \caption{Convergence of LEs and forward-backward evolution to calculate CLVs for $t_{\text{step}} = 0.3$.}
\label{fig:CLValgosketch}
\end{figure}

To compute the LEs of the flow under consideration
, we evolve 6 orthonormal perturbations \(\{\bm{\delta \mathbf{u}} \}_{i=1 \dots 6}\) in the tangent space. The general perturbations $\bm{\delta \mathbf{u}}$ can be expressed as the Fourier integral in the spanwise direction, 
\begin{equation}
\bm{\delta \mathbf{u}}(x,y,z,t)=\int_{-\infty}^{+\infty} \bm{\widehat {\delta {\mathbf{u}}}}(x,y,t; \beta) e^{\iota \beta z} dz
\end{equation}
and similarly for $\delta p$. In the present work, we take $\beta=0$, i.e.\ we consider only 2D perturbations. The perturbations are orthonormalized periodically using Gram-Schmidt QR decomposition. Two segment lengths are chosen \(t_{\text{step}} = 0.3, 10\). The convergence of the LEs with the integration time for $t_{\text{step}} = 0.3$ is shown in figure \ref{fig:CLValgosketch}. 

The unstable manifold consists of two positive LEs, which indicates that, on average, perturbations grow exponentially along two directions at any point on the attractor. The leading LE is \(\lambda_1 \approx 0.1\). One LE is equal to 0 (neutral manifold) which indicates the absence of growth or decay of perturbations along the trajectory in phase space \citep{Towards_chaotic_adjoint_LES_blonigan}. Furthermore, the system is non-degenerate as all LEs are distinct, \cite{ginelli_covariant_2013}. 

\begin{figure}[ht!]
\centerline{\includegraphics[width=0.7\linewidth]{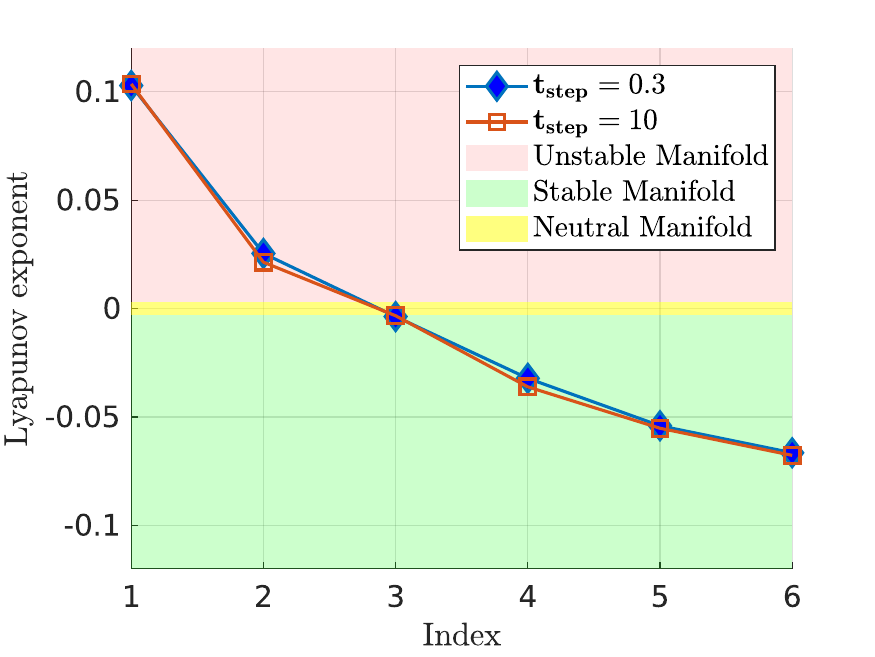}}
\caption{Lyapunov spectrum of the flow.}
\label{LEsfig}
\end{figure}

The LEs were also computed for \(t_{\text{step}} = 10\), which is approximately one Lyapunov time unit, $\frac{1}{\lambda_1}$. The LE spectra for the two values of \(t_{\text{step}}\) are shown in \ref{LEsfig}. The results are almost identical, confirming the robustness of the calculation.

\subsection{Covariant Lyapunov Vectors} \label{sec:CLVs}

It was shown that LEs converge to constant values when averaged over a long time window. The corresponding directions of growth/decay are known as Covariant Lyapunov Vectors (CLVs), see  \cite{ginelli_characterizing_2007, ginelli_covariant_2013}. CLVs provide a geometric characterization of the tangent space. They are unit-norm vectors that span the Gram-Schmidt (GS) subspace and represent the expanding, contracting, and neutral directions at each point of the attractor. These directions correspond to the positive, negative, and zero LEs, respectively.

The concept of CLVs was first introduced by \cite{Oseledets1968} and later formalized as tangent directions of invariant manifolds by \cite{Ruelle1979}. However, for decades there was no effective algorithms for their computation. \cite{ginelli_characterizing_2007} proposed a stable algorithm to extract CLVs, leveraging the geometric understanding of the tangent space of  \cite{Ruelle1979}. The algorithm is based on a forward-backward evolution of the tangent space. A brief overview is presented in Algorithm \ref{Ginelli_algo} below; the  detailed methodology of \cite{ginelli_covariant_2013} is presented in  \ref{CLVcalcMethodappendix}. 

\begin{algorithm}[ht!]
\caption{Calculation of  Covariant Lyapunov Vectors  \citep{ginelli_covariant_2013}}\label{alg:clv}
\begin{algorithmic}[1]
    \State Given a generic initial condition \(\mathbf{u}_0\), integrate \( \dfrac{d \mathbf{u}}{dt} = \mathbf{f}(\mathbf{u})\) until the trajectory has reached the ergodic attractor.
    \vspace{2mm}
    \Statex \underline{\textbf{Forward Transient}}
    \State Initialize orthonormal perturbations \(\left\{\bm{\delta \mathbf{u}}_j\right\}_{j=1 \dots m}\) and reset time to \(t = 0\).
    \State Divide the total time window $[0,T]$ into $K$ time steps, each of length \(t_{\text{step}}\).
    \State Evolve phase space dynamics \(\mathbf{u}(T)=\mathcal{L}^{T}(\mathbf{u}_0) \) and tangent space dynamics ${\bm{\delta \mathbf{u}}}_j \to D \mathcal{L}^{t_{\text{step}}}(\mathbf{u}) \bm{\delta \mathbf{u}}_j$ until the subspace 
    $\left\{\bm{\delta \mathbf{u}}_j\right\}_{j=1 \dots m}$ converges.
    \vspace{2mm}
    \Statex \underline{\textbf{Forward Dynamics}}
    \State Evolve the phase and tangent space dynamics further until \(t = T + t_{\text{step}}(k_1 + k_2)\).
    \State Perform Gram-Schmidt QR decomposition and store the Gram-Schmidt vectors \(\mathbf{G}(\{\bm{\delta \mathbf{u}}_j\}_{j=1 \dots m})\) and the upper-triangular \(R\) matrices.
    \vspace{2mm}
    \Statex \underline{\textbf{Backward Transient}}
    \State Evolve a generic non-singular lower triangular matrix \(\tilde{\mathbf{C}}_{T + t_{\text{step}}(k_1 + k_2)}\) backwards for time \(k_2 t_{\text{step}}\) to converge the CLV expansion coefficients (that are stored in matrix \(\tilde{\mathbf{C}}\) ).
    \State Normalize the columns of \(\tilde{\mathbf{C}}\) every $t_{\text{step}}$
    \vspace{2mm}
    \Statex \underline{\textbf{Backward Dynamics}}
    \State Evolve the converged CLV expansion coefficients \(\tilde{\mathbf{C}}_{T + k_1 t_{\text{step}}}\) further backwards. 
    \vspace{2mm}
    \Statex \underline{\textbf{Calculate the CLVs}}
    \State With the converged GS-vector subspace \(\mathbf{G}_{T}\) and the true CLV expansion coefficients \(\mathbf{C}_{T}\), calculate the CLVs (\(\mathbf{V}\)) from \(T\) to \(T + k_1 t_{\text{step}}\).
\end{algorithmic}
\label{Ginelli_algo}
\end{algorithm}

The 4 phases of the algorithm are shown in figure \ref{fig:CLValgosketch}. The algorithm begins with the forward transient phase (in our case \(t \in [0, 2100]\)), in which the dynamical system and  six (initially random) orthonormal perturbations are evolved until the tangent subspace and the LE spectrum converge. In the forward dynamics phase ($t \in [2100, 4500]$), the orthonormal Gram-Schmidt vectors  $\mathbf{G}_n = \left(\mathbf{g}_n^{(1)}\mid\mathbf{g}_n^{(2)}\mid \ldots \mid \mathbf{g}_n^{(6)}\right)$ and the $R_n$ matrices (obtained from the QR decomposition) are stored for every time segment, $n$. In the backward transient phase ($t \in [5400, 4500]$) the CLV expansion coefficient matrix \(\mathbf{C}_n\) is evolved backwards. To determine the length of this phase, \(k_2 t_{\text{step}}\) was increased and the angles between the CLVs in the backward dynamics phase were monitored. At \(k_2 t_{\text{step}} = 150\), the angles between the CLVs stabilized, indicating convergence.  In the final backward dynamics phase, the recorded GS matrices \(\mathbf{G}_n\) are multiplied with the expansion coefficient matrices \(\mathbf{C}_n\) over the middle time horizon \(t \in [2100, 4500]\) to generate 8000 samples of six CLVs (see equation \eqref{CLVeq1} in \ref{CLVcalcMethodappendix}). 

Each CLV \(\mathbf{V}_i\) has dimension \(2N\) and can be expressed as \(\mathbf{V}_i = [V_i^u, V_i^v ]\), where superscripts \(u\) and \(v\) denote the \(x\) and \(y\) velocity components, respectively. The magnitude of the $i$-th CLV, \( |\mathbf{V}_i| \), is defined as the Euclidean norm, i.e. \(|\mathbf{V}_i|=\sqrt{(V_i^u)^2 + (V_i^v)^2}\).

\begin{figure}[ht!]
    \centering
    \includegraphics[width=1\linewidth]{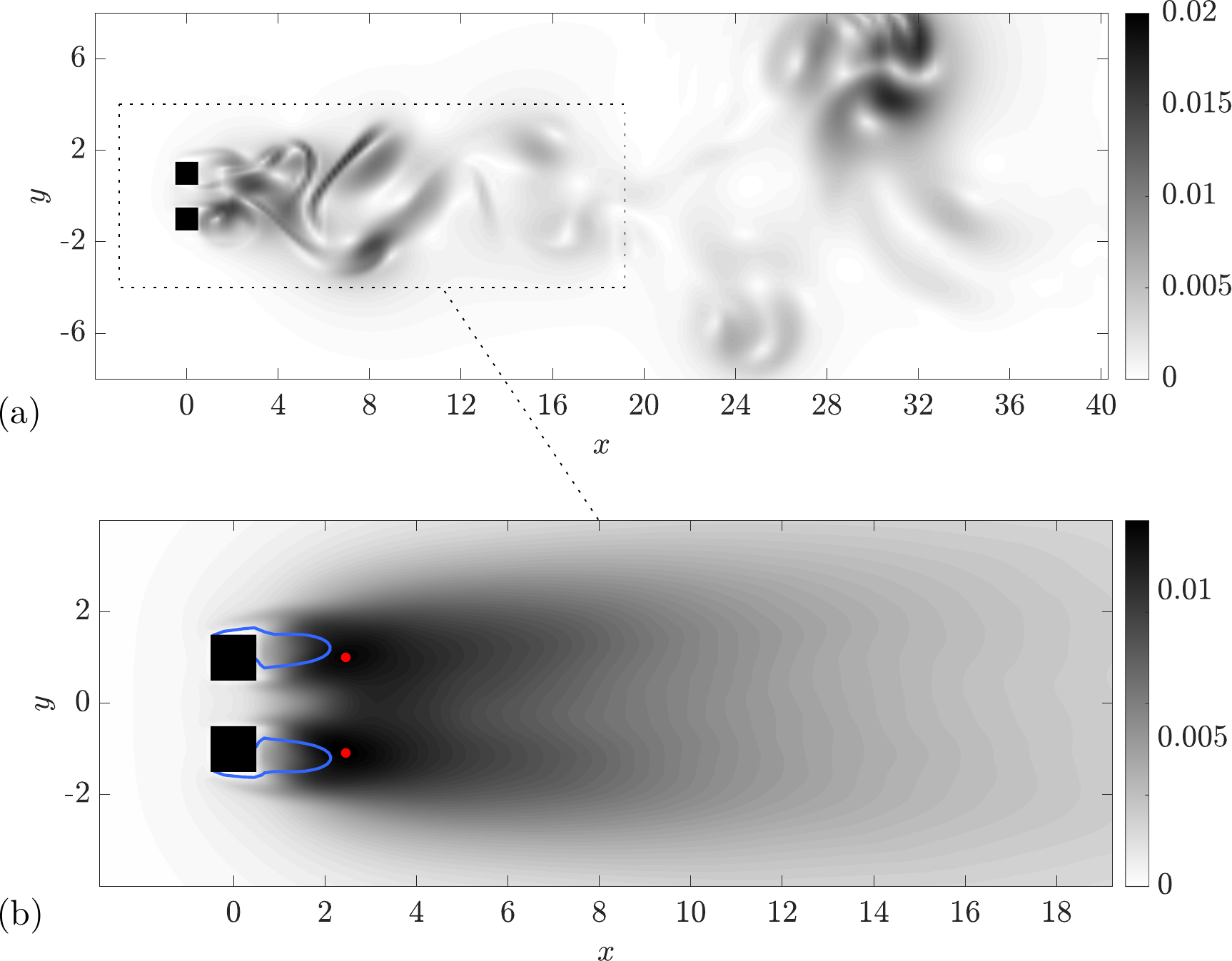}
    \caption{Contour plots of the magnitude of the leading CLV \(|\mathbf{V}_1|\). (a) Instantaneous \(|\mathbf{V}_1|\) and (b) time-averaged \(|\mathbf{V}_1|\). The blue isolines mark the recirculation regions behind the cylinders, and the red dots indicate the locations of peak magnitude. For an animation of  \(|\mathbf{V}_1|\) contours, see supplementary video.}
    \label{CLV1meanandinst}
\end{figure}

Figure~\ref{CLV1meanandinst} shows contours of the instantaneous and time-averaged magnitude of the leading unstable CLV, $|\mathbf{V}_1|$ (note that the two plots have different streamwise and cross-stream extents). The instantaneous contours (top panel) reveal that $|\mathbf{V}_1|$ takes the form of coherent structures close to the cylinders but the values are decaying for $x>8$. Occasional high values are detected much further downstream and away from the centerline (in the time instant shown they are in $x\in [28,34]$). These high values are concentrated around vortices that have formed close to the cylinders and are ejected away from the centerline by the sweeping motion of the flapping jet. They are then captured by the freestream and transported for long distances. The time-averaged contours are symmetric (as expected) and peak above and below the gap centerline at $x =2.5$, slightly downstream of the recirculation zone. The large  footprint extends up to $x \approx 8$ and then the values decay. The high values that were detected far downstream in the instantaneous plot are occasional and do not affect the time-average. 

\begin{figure}[ht!]
    \centering
    \includegraphics[width=1\linewidth]{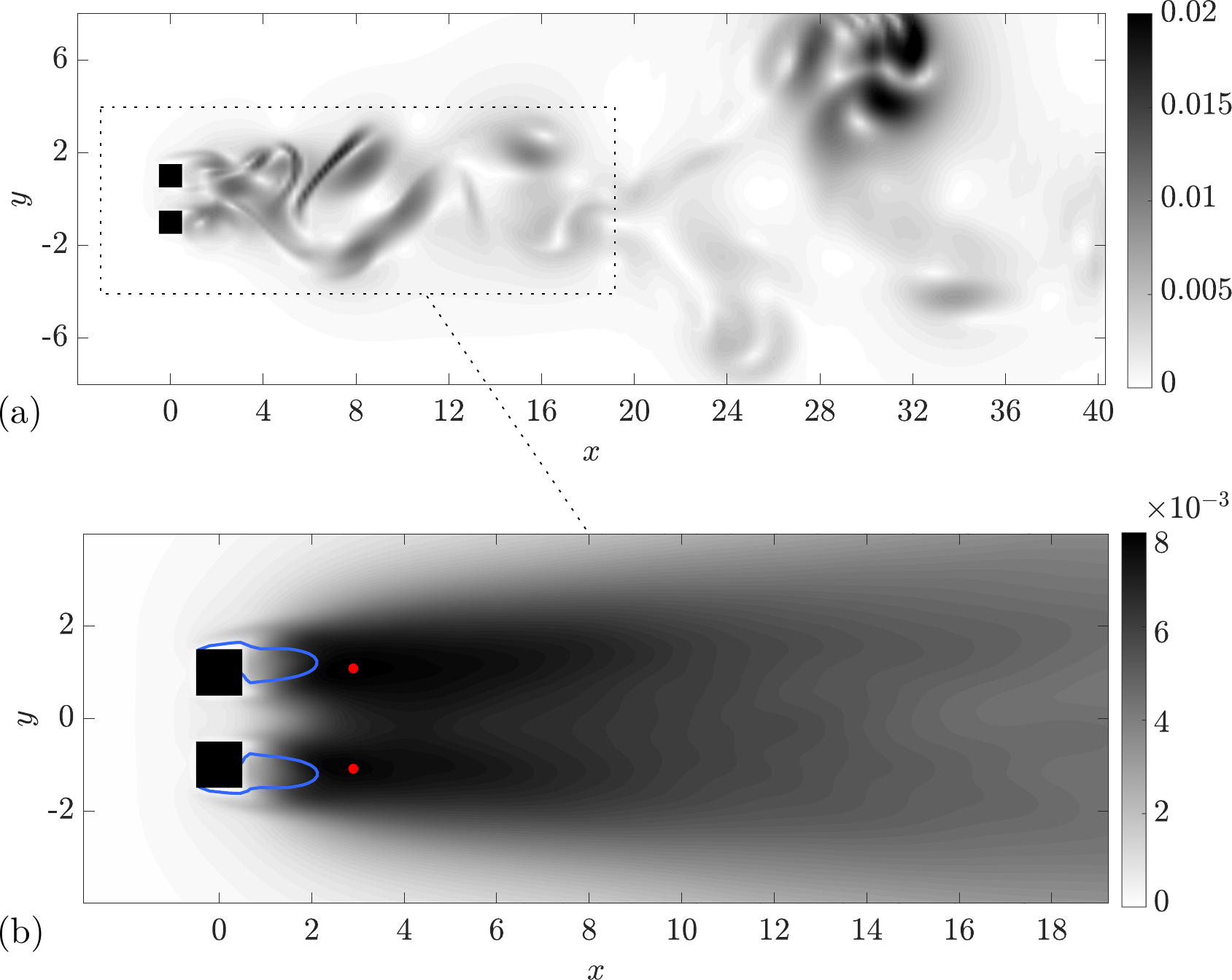}
    \caption{Contour plot of the second unstable CLV: (a) Instantaneous \(|\mathbf{V}_2|\) and (b) Mean \(|\mathbf{V}_2|\). The blue isolines mark the recirculation regions behind the cylinders, and the red dots indicate the locations of peak magnitude.  For an animation of  \(|\mathbf{V}_2|\) contours, see supplementary video.}
    \label{CLV2meanandinst}
\end{figure}

Figure~\ref{CLV2meanandinst} shows contours of the magnitude of the second unstable CLV, $|\mathbf{V}_2|$. The instantaneous magnitude (visualised at the same time instant as that of $\mathbf{V_1}$) appears slightly weaker in the near wake region ($x \leq 4$) achieving higher intensity further downstream. Again a region of high values appears further downstream and away from the centerline. The time-average peaks at $x = 2.9$ but maintains high values until approximately $x = 14-16$.  This observation suggests that $\mathbf{V}_2$ has presence further downstream compared to $\mathbf{V}_1$.  

It is clear that there is a link between the structures of the unstable CLVs and the flow characteristics analysed earlier. This connection will be elucidated in detail in section \ref{sec:SPOD_of_unstable_CLVs} below. The direction of the neutral CLV $\mathbf{V}_3$ is along the system trajectory in phase space i.e. $\frac{d \mathbf{u}}{dt}$ and is not shown. We therefore proceed with examining one of the CLVs of the stable subspace.

\begin{figure}[ht!]
    \centering
    \includegraphics[width=1\linewidth]{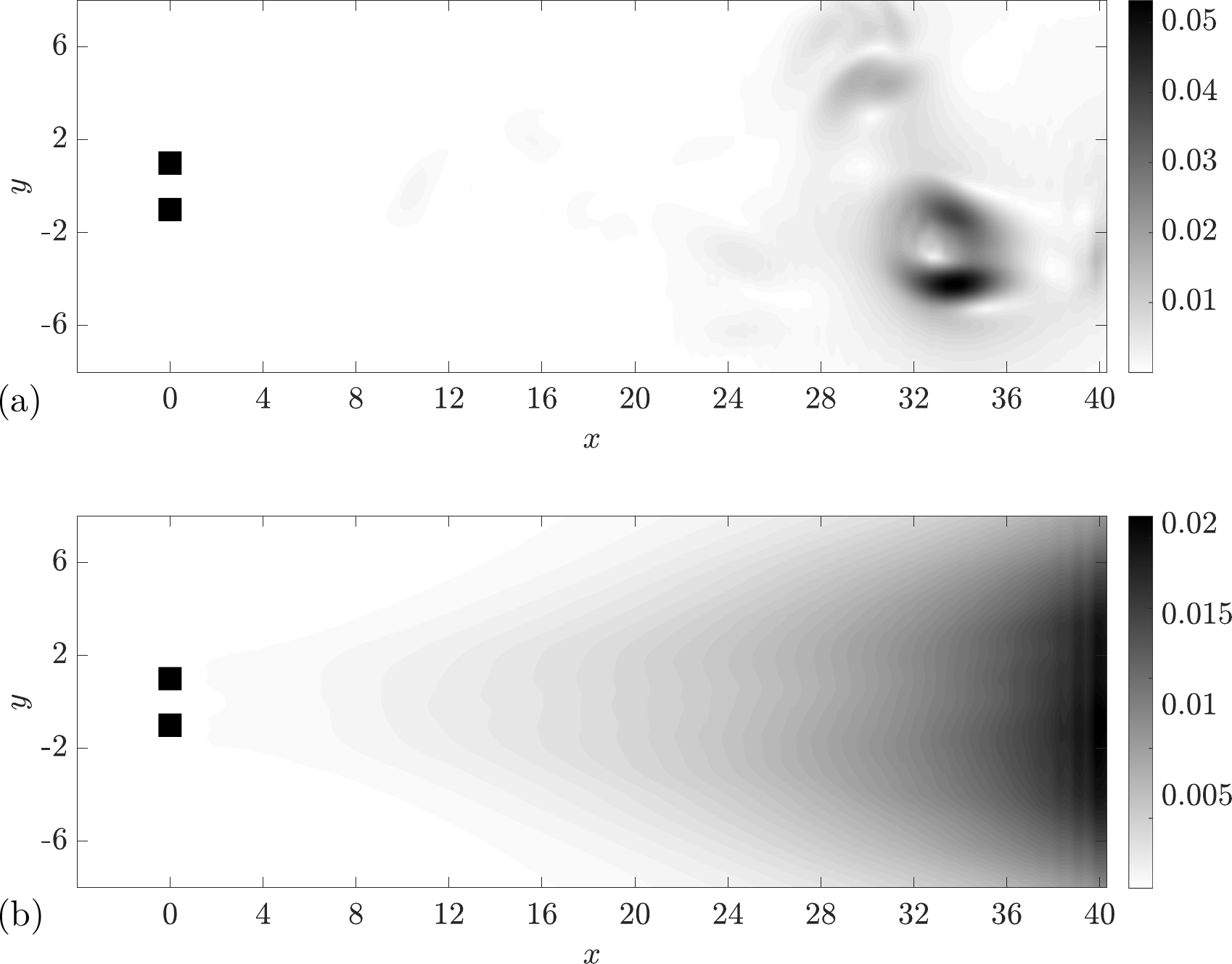}
    \caption{Contour plot of the stable CLV: (a) Instantaneous \(|\mathbf{V}_6|\) and (b) Mean \(|\mathbf{V}_6|\) (Bottom).}
    \label{CLV6meanandinst}
\end{figure}

Figure \ref{CLV6meanandinst} depicts contours of the magnitude of the stable CLV, $|\mathbf{V}_6|$. The instantaneous contour (visualised at the same time instant as that of $\mathbf{V_1}$ and $\mathbf{V_2}$) reveals negligible footprint until approximately $x \approx 25$, with the intensity peaking further downstream, at $x \approx 35$. The time-averaged contour shows that the strongest activity of this CLV is concentrated near the outlet boundary, where dissipation overwhelmingly dominates production, consistent with the negative Lyapunov exponent associated with $\mathbf{V}_6$. These trends align with the findings of \citet{ni_hyperbolicity_2019} who investigated the flow around a cylinder and reported a systematic downstream shift of the spatial CLV footprints as the LE values decrease.

\subsection{Assessment of Hyperbolicity}

In uniformly hyperbolic systems, the tangent space can be decomposed distinctly into unstable ($\mathbf{E}_{\mathbf{u}}^u$), stable ($\mathbf{E}_{\mathbf{u}}^s$), and neutral ($\mathbf{E}_{\mathbf{u}}^0$) subspaces,
\begin{equation}
\mathcal{T}_{\mathbf{u}} \mathcal{M}=\mathbf{E}_{\mathbf{u}}^u \oplus \mathbf{E}_{\mathbf{u}}^s \oplus \mathbf{E}_{\mathbf{u}}^0,
\end{equation}
at any point on the attractor. These subspaces are spanned by the CLVs corresponding to positive, negative and zero LEs respectively. For a system to be classified as uniformly hyperbolic, the three subspaces must be bounded away from each other, i.e.\ they should not align at any point on the attractor. For the flow examined, $\mathbf{V_1}$ and $\mathbf{V_2}$ span $\mathbf{E}_{\mathbf{u}}^u$, $\mathbf{V_3}$ spans $\mathbf{E}_{\mathbf{u}}^0$, and $\mathbf{V_4}$, $\mathbf{V_5}$, and $\mathbf{V_6}$ span $\mathbf{E}_{\mathbf{u}}^s$. Note that we have determined only three stable CLVs, therefore our understanding of the stable subspace is limited. 

The angle \( \phi^{i,j} \) between \( \mathbf{V}_i \) and \( \mathbf{V}_j \), is calculated as:
\begin{equation}
    \phi^{i,j} = \cos^{-1} \left( |\mathbf{V}_i \cdot \mathbf{V}_j| \right)
\end{equation}
where \( \mathbf{V}_i \cdot \mathbf{V}_j \) is the dot product (defined as $\mathbf{V}_i \cdot \mathbf{V}_j$, and \( |\cdot| \) is the absolute value. The above expression provides an angle between $0^{\circ}$ and $90^{\circ}$ degrees  \cite{ginelli_characterizing_2007, xuandpaul2016}. If $\phi^{i,j}=0$ the two CLVs are parallel to each other, while if $\phi^{i,j}=90$ they are orthogonal. 

\begin{figure}[ht!]
    \centering
    \includegraphics[width=1\linewidth]{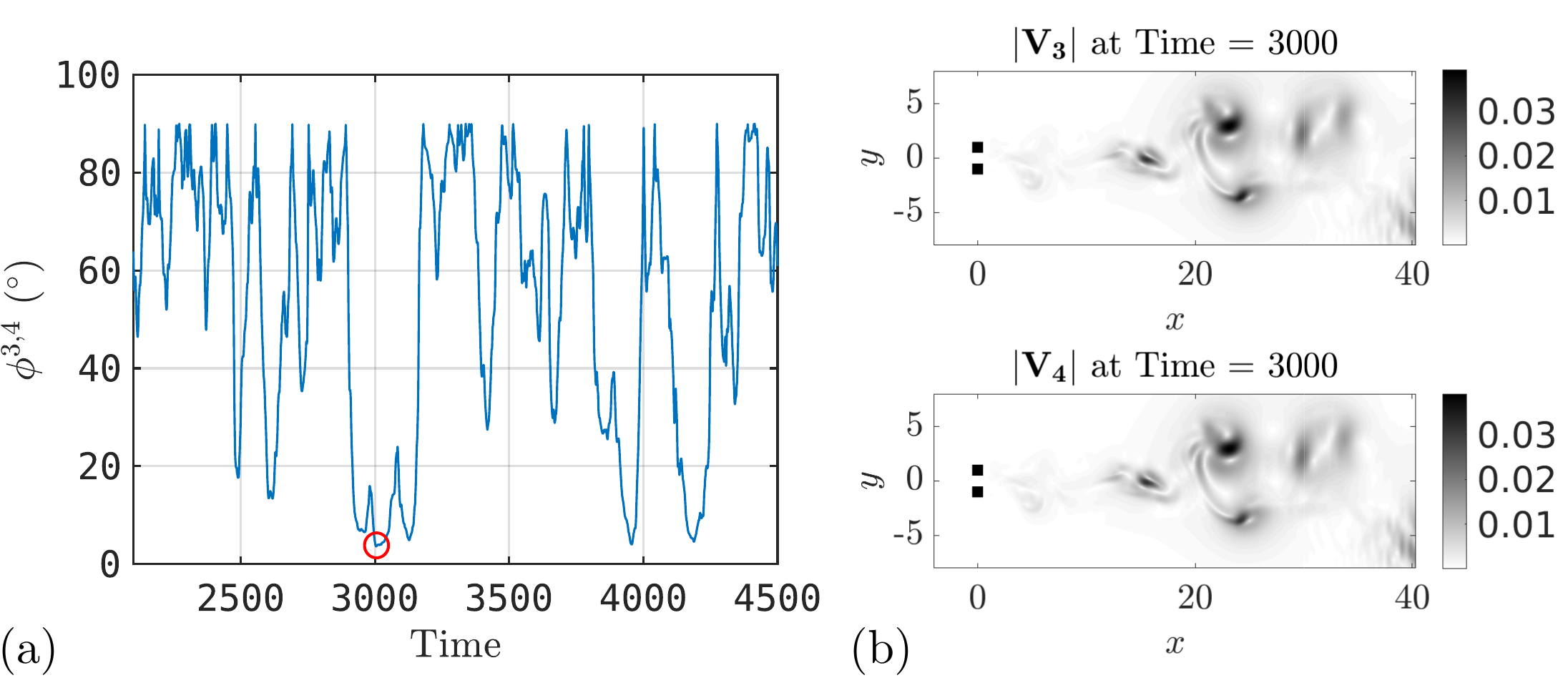}
    \caption{Variation of the angle between $\mathbf{V_3}$ and $\mathbf{V_4}$ over time (left), contour plots of instantaneous $|\mathbf{V_3}|$ and $|\mathbf{V_4}|$ at $ = 3000$ (right).}
    \label{visualviolationinHyperbolicity}
\end{figure}

The angles $\phi^{i,j}$ are computed from 8000 samples taken every $t_{\text{step}} = 0.3$ time units in the time window $[2100, 4500]$, see figure \ref{fig:CLValgosketch}. Figure \ref{visualviolationinHyperbolicity} illustrates the angle between $\mathbf{V_3}$ and $\mathbf{V_4}$ in this time window. Notice that $\phi^{3,4}$ approaches zero at $t = 3000$. Contours of the magnitude of the two CLVs at this time instant are shown in the right panel. The plots are most identical confirming that the CLVs are parallel, which indicates violation of hyperbolicity at this time instant.

\begin{figure}[ht!]
    \centering
    \includegraphics[width=1\linewidth]{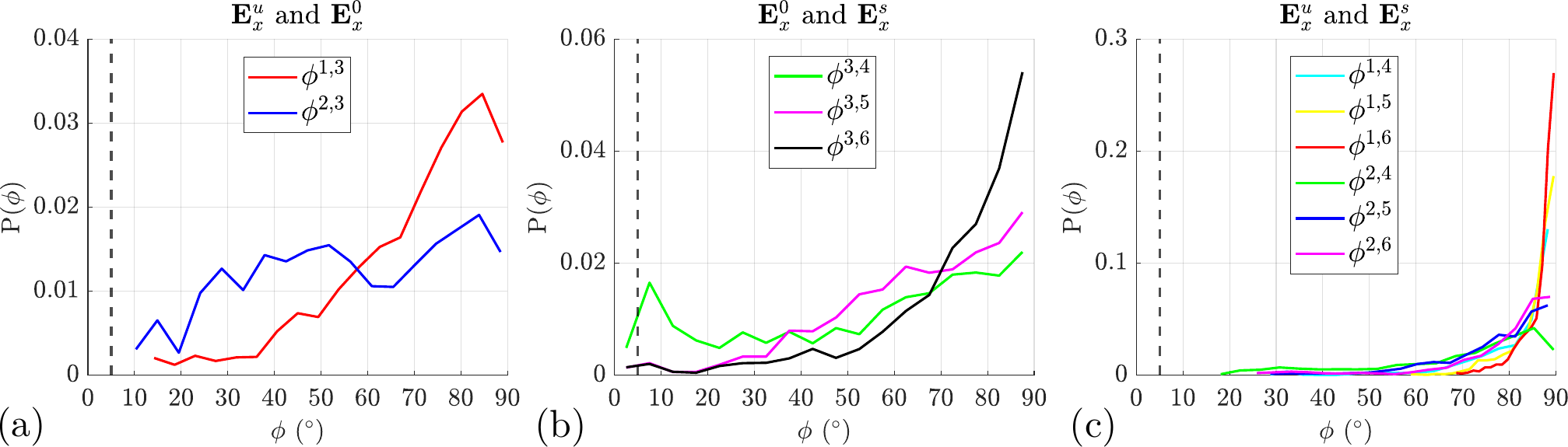}
    \caption{Probability density functions (PDFs) of angles between CLVs spanning (a) the unstable and neutral subspaces , (b) the neutral and stable subspaces, and (c) the unstable and stable subspaces. The vertical dashed lines indicate the angle $\phi=5^{\circ}$.}
    \label{fig:CLVangles}
\end{figure}

Figure \ref{fig:CLVangles} shows the probability density functions (PDFs) of the angles between pairs of CLVs that belong to different subspaces. The domain  $[0^{\circ},90^{\circ}]$ is segmented into 18 bins (bin width $5^{\circ}$). We consider that samples located in the first bin $[0^{\circ}, 5^{\circ}]$ signify instances of hyperbolicity violation. Notice that CLVs that belong to different subspaces but nearby indices $i$ and $j$, have a higher probability of being parallel (or nearly-parallel) to each other. For example the PDFs of $\phi^{2,3}$, $\phi^{3,4}$, and $\phi^{3,5}$ approach (or have samples) within the first bin. 

On the other hand, when the indices are far apart the CLVs have higher probability of orthogonality, see for example the PDFs of  $\phi^{1,3}$, $\phi^{3,6}$, $\phi^{1,5}$ and especially $\phi^{1,6}$ which has a sharp spike at $90^{\circ}$. The orthogonality can be explained by the non-overlapping footprints of the CLVs. For example, $\mathbf{V_1}$ and $\mathbf{V_6}$ peak in the near wake and far-wake regions respectively, see figures \ref{CLV1meanandinst} and \ref{CLV6meanandinst}, thus the overlap is minimal and the CLVs are almost orthogonal. 

It was found that hyperbolicity is violated  in $2.4\%$ of $\phi^{3,4}$ samples, $0.67\%$ of the $\phi^{3,5}$ samples and $0.69\%$ of the $\phi^{3,6}$ samples. These angle combinations correspond to CLVs that belong to neutral and stable subspaces, and indicate that hyperbolicity is mainly violated due to the alignment of these two subspaces. These observations are consistent with the findings of \cite{ni_hyperbolicity_2019} and \cite{xuandpaul2016} who studied hyperbolicity for the flow around a circular cylinder and for Rayleigh-Bénard convection respectively.

\section{Spectral POD of the unstable CLVs} \label{sec:SPOD_of_unstable_CLVs}

We now turn our attention to the two unstable CLVs, \( \mathbf{V_1} \) and \( \mathbf{V_2} \), and analyze their spatiotemporal characteristics using Spectral POD. This method can extract the dominant structures of the CLVs at each frequency and facilitates comparison with the eigenmodes obtained with GLSA.  

We assemble matrices \( \tilde{\mathbf{Y}}_1 \) and \( \tilde{\mathbf{Y}}_2 \) by stacking column by column $8000$ snapshots of \( \mathbf{V_1} \) and \( \mathbf{V_2} \) respectively. Zero-mean stochastic ensembles are generated by subtracting the time-averages, resulting in \( \mathbf{Y}_1 = \tilde{\mathbf{Y}}_1 - \overline{\mathbf{Y}}_1 \) and \( \mathbf{Y}_2 = \tilde{\mathbf{Y}}_2 - \overline{\mathbf{Y}}_2 \). The matrices are divided into \( N_b = 61\) blocks each containing \( N_f = 256 \) snapshots with an overlap of 50\%. 

\subsection{\texorpdfstring{Analysis of the most unstable CLV, $\mathbf{V}_1$}{Analysis of the second unstable CLV, V1}}

\begin{figure}[ht!]
    \centering
    \includegraphics[width=0.75\linewidth]{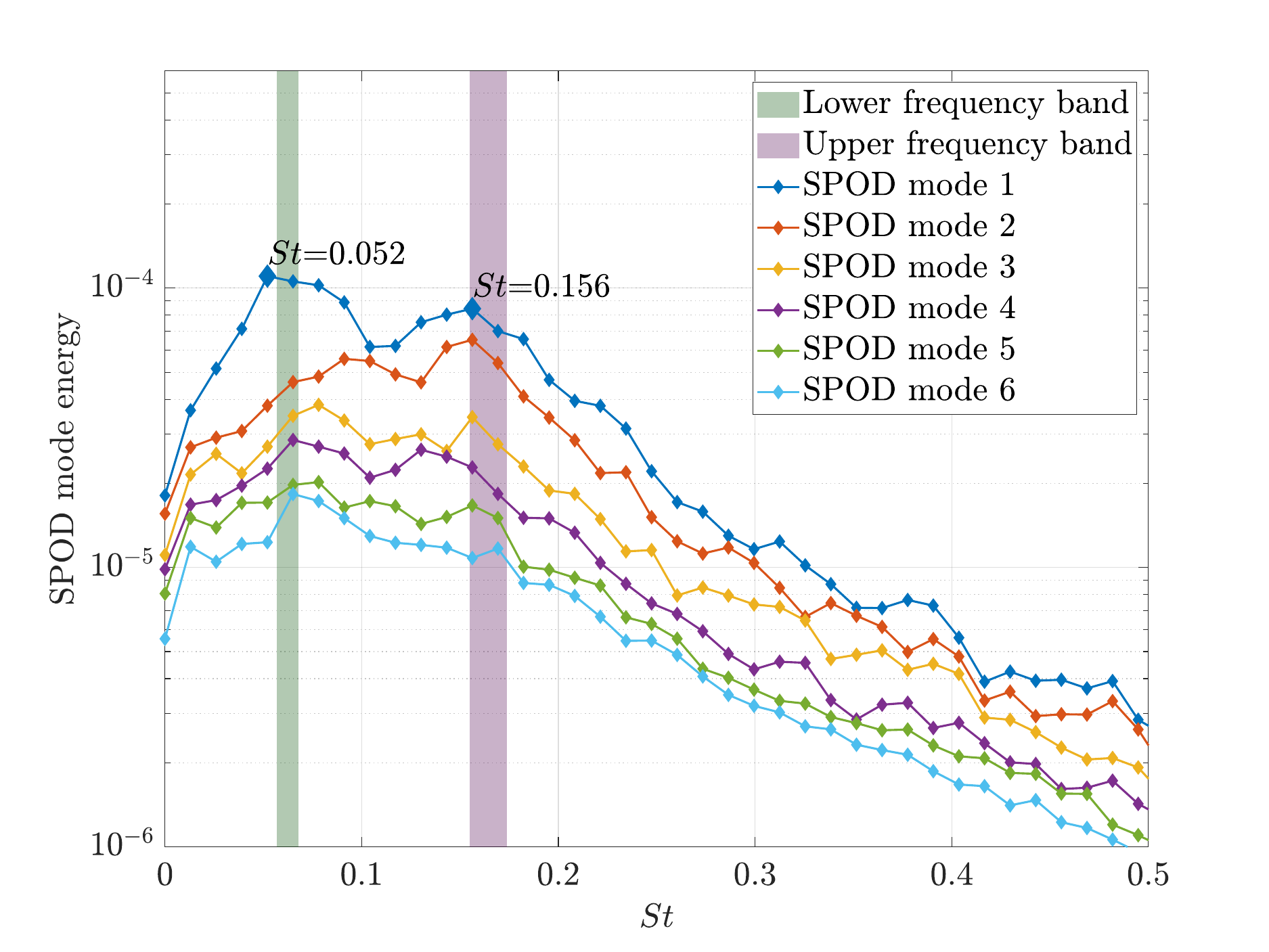}
    \caption{Energy spectrum of the first six SPOD modes of $\mathbf{V_1}$.}
    \label{SPOD_CLV1_energyspectrum}
\end{figure}

The energy spectrum of the first six SPOD modes of \( \mathbf{V}_1 \) is shown in figure ~\ref{SPOD_CLV1_energyspectrum}. Two distinct peaks are observed that lie within the lower and upper frequency bands. The energy separation between the first and second SPOD modes is strongest in the lower frequency band, as in the flow field modes. 

\begin{figure}[ht!]
    \centering
    \includegraphics[width=\linewidth]{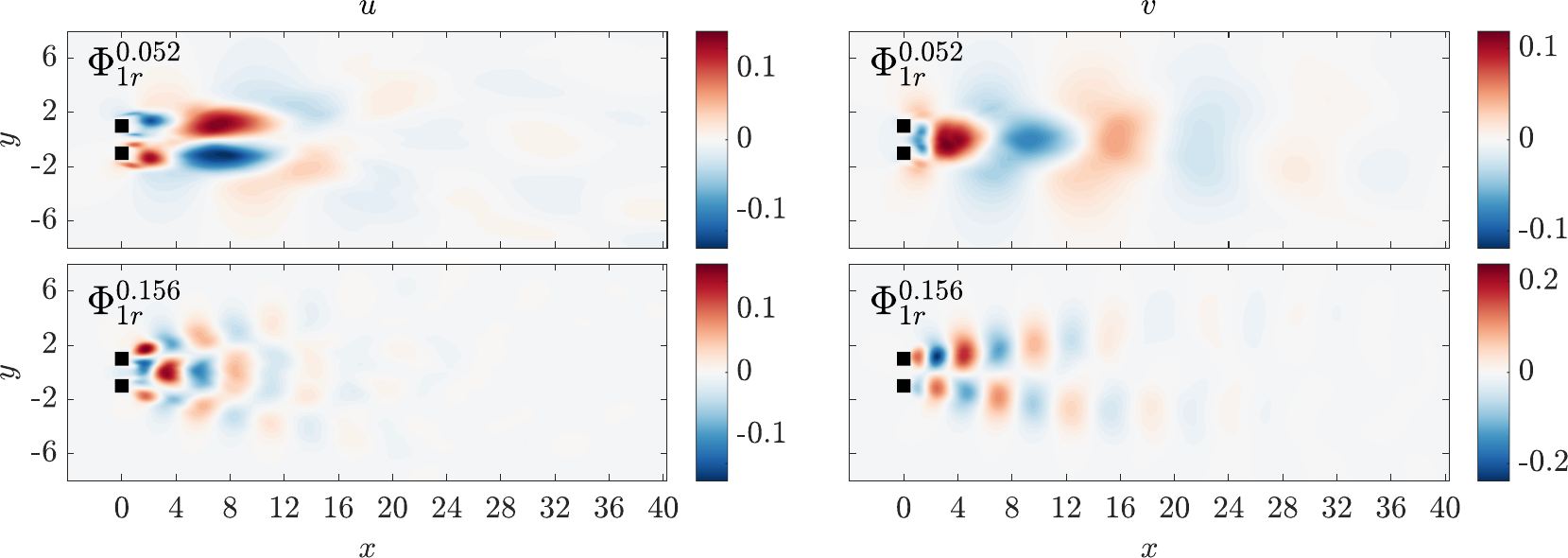}
    \caption{Contour plots of streamwise velocity (left column) and cross-stream velocity (right column) of  the real part of SPOD mode 1 of $\mathbf{V_1}$ at \( St = 0.052 \) (top row) and \( St = 0.156 \) (bottom row) i.e. $\Phi_{1r}^{0.052}$ and $\Phi_{1r}^{0.156}$.}
    \label{SPOD_CLV1_contours}
\end{figure}

\begin{figure}[ht!]
    \centering
    \includegraphics[width=\linewidth]{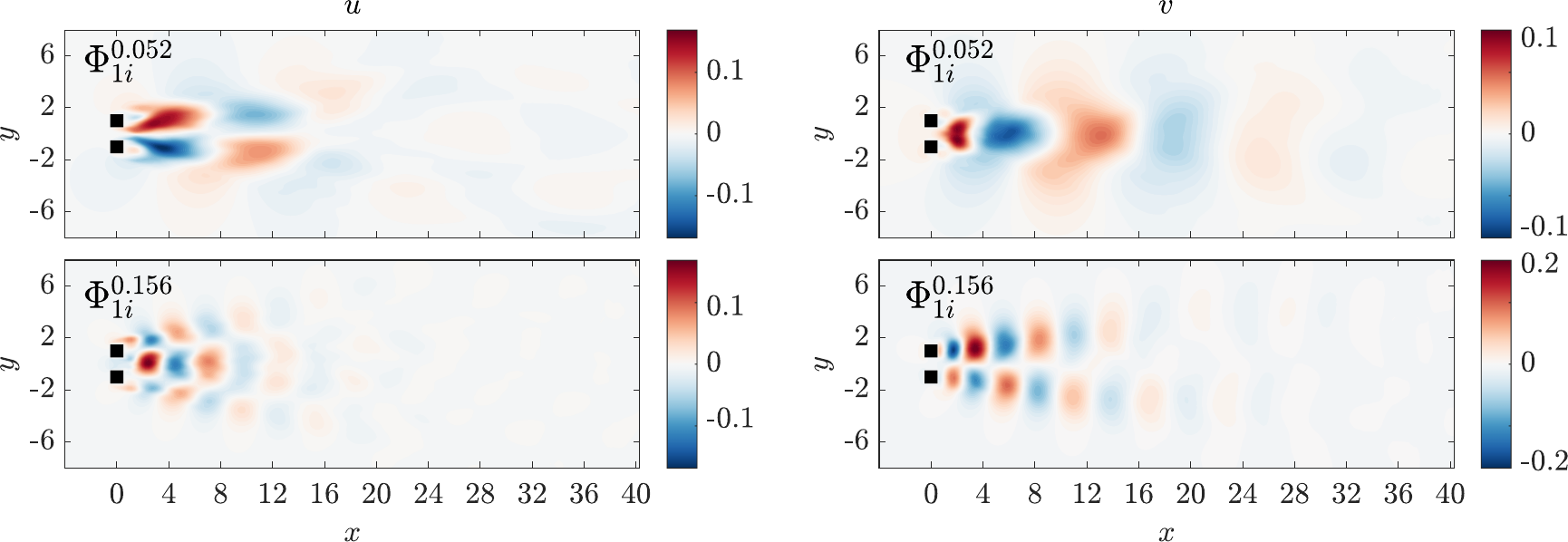}
    \caption{Contour plots of streamwise velocity (left column) and cross-stream velocity (right column) of  the imaginary part of SPOD mode 1 of $\mathbf{V_1}$ at \( St = 0.06 \) and \( St = 0.156 \) i.e. $\Phi_{1i}^{0.06}$ and $\Phi_{1i}^{0.156}$.}
    \label{SPOD_CLV1_contours_imag}
\end{figure}

The real and imaginary parts of the dominant SPOD mode at these peaks are shown in figures \ref{SPOD_CLV1_contours} and \ref{SPOD_CLV1_contours_imag} respectively. In each figure, contour plots of streamwise velocity ($u$) and cross-stream velocity ($v$) are shown on the left and right panels respectively. Note that the imaginary part is spatially shifted with respect to the real part, consistent with a propagating structure beating at the frequency shown. At the lower peak $St = 0.052$, which is close to the flapping frequency of the flow (see figure \ref{fig:SPOD_nonlinear}), the SPOD mode exhibits structures characteristic of the flapping jet. The $v$ contours of both the real and imaginary parts show alternating positive and negative peaks along the gap centerline that decay downstream. The alternating signs in the near wake are consistent with the jet-flapping up and down motions. These features demonstrate that the low frequency band of  \( \mathbf{V}_1 \) captures coherent structures in the near-wake that peak in the gap centerline. In the $v$ contour of the higher band at $St = 0.156$, a train of small-scale structures is observed to shed from the cylinders and propagate downstream while expanding in the cross-stream direction. These structures correspond to the vortex shedding behind the cylinders and also closely resemble those of the SPOD mode 1 of the flow in the same frequency band. These observations confirm that the dominant SPOD mode of \( |\mathbf{V}_1| \) accurately captures the main instability features of the flow in both the lower and upper frequency bands. However, it does not contain any signature of the subharmonic.

\subsection{\texorpdfstring{Analysis of the second unstable CLV, $\mathbf{V}_2$}{Analysis of the second unstable CLV, V2}}

\begin{figure}[ht!]
    \centering
    \includegraphics[width=0.75\linewidth]{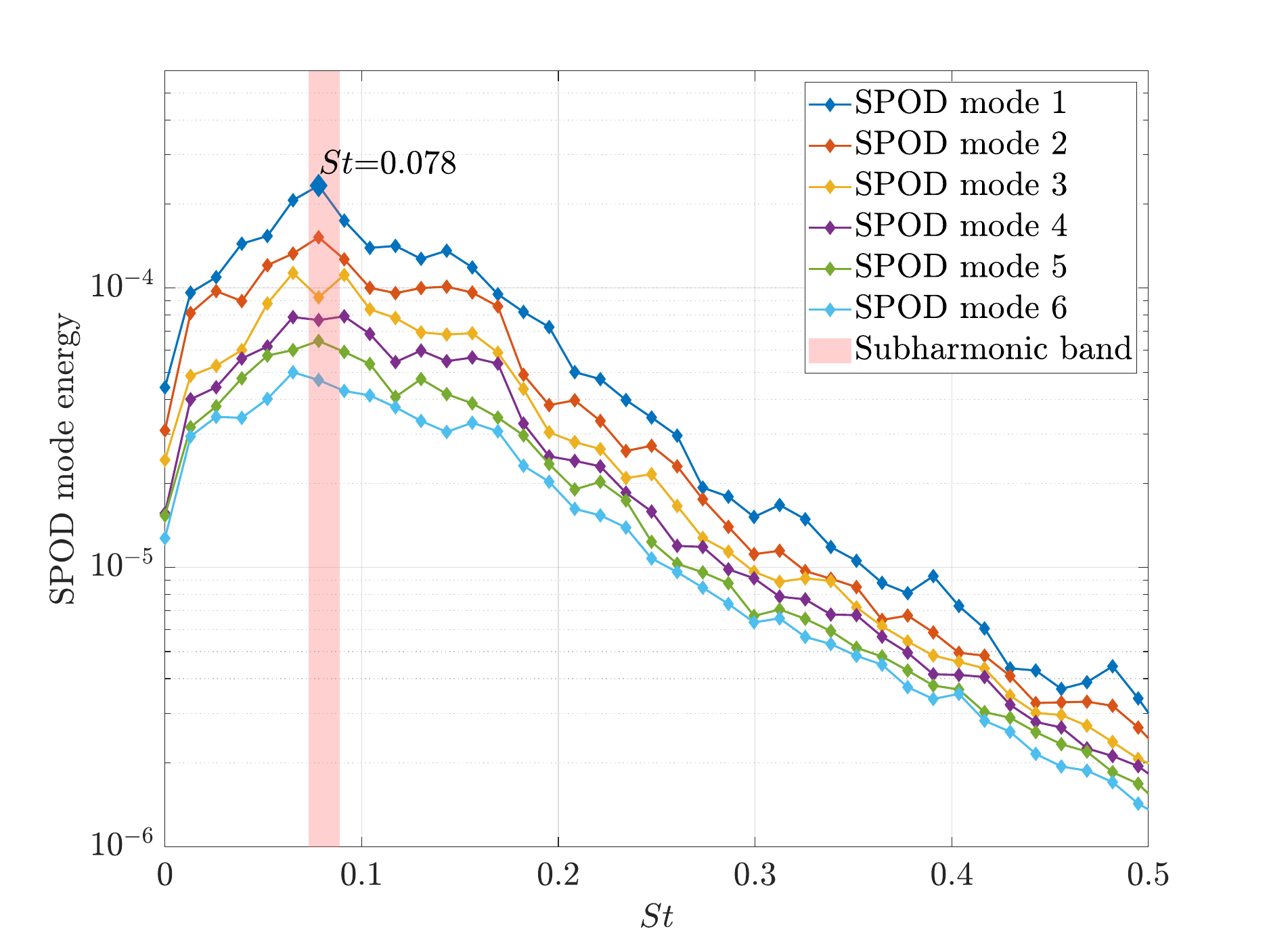}
    \caption{Energy spectrum of the first six SPOD modes of \( \mathbf{V_2} \).}
    \label{SPOD_CLV2_energyspectrum}
\end{figure}

The energy spectrum of the first six SPOD modes of the second unstable CLV, \( \mathbf{V}_2 \), is shown in figure~\ref{SPOD_CLV2_energyspectrum}. While the energy of $\mathbf{V}_1$ is spread across two frequency bands, the dominant SPOD mode of $\mathbf{V}_2$ exhibits a single prominent peak at $St = 0.078$, followed by a rapid decay. At this frequency, the energy of SPOD mode 1 is approximately three times that of mode 2, underscoring its importance. This frequency lies in the subharmonic band and is precisely half of the primary vortex shedding frequency $St = 0.156$,  indicating that $\mathbf{V}_2$ captures the subharmonic instabilities resulting from interactions between paired vortices. 

\begin{figure}[ht!]
    \centering
    \includegraphics[width=1\linewidth]{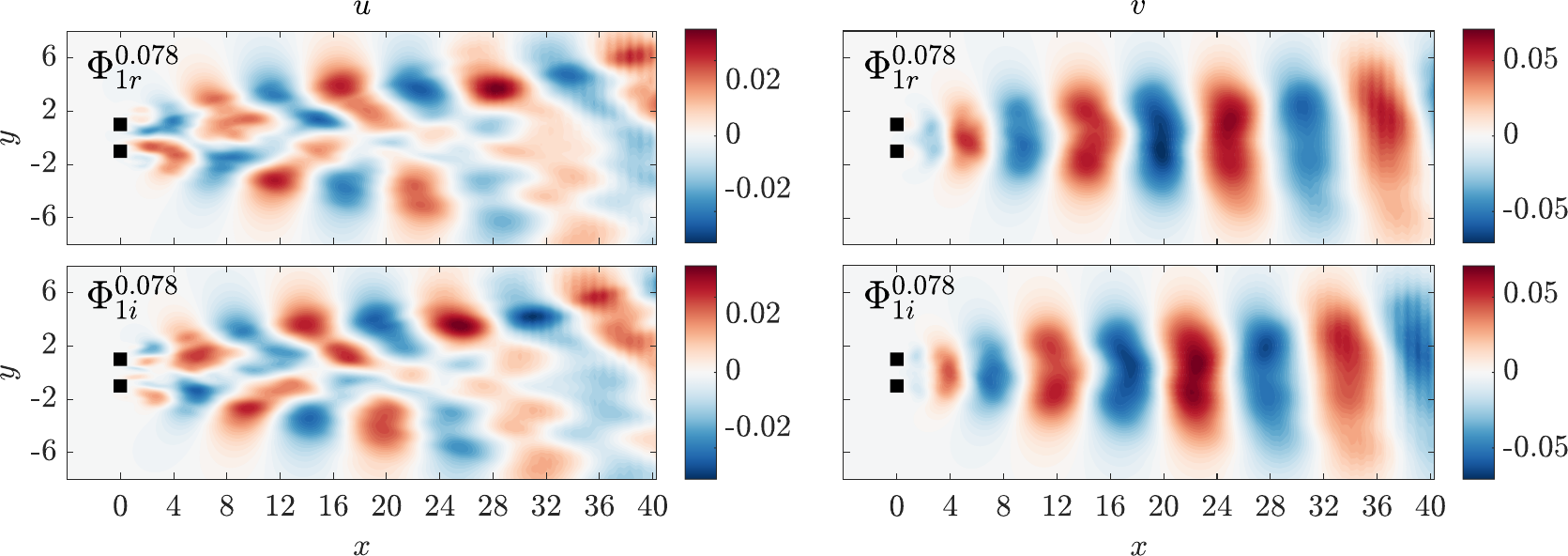}
    \caption{Contour plots of streamwise velocity (left column) and cross-stream velocity (right column) of  the real part of SPOD mode 1 of \( \mathbf{V_2} \) at \( St = 0.078 \) i.e. $\Phi_{1r}^{0.078}$.}
    \label{SPOD_CLV2_contours}
\end{figure}

Contour plots of the real part of the streamwise (\( u \)) and cross-stream (\( v \)) velocity components of SPOD mode 1 at \( St = 0.078 \) are shown in figure~ \ref{SPOD_CLV2_contours}. Large alternating red and blue bands appear in the \( v \) contours in the mid-to-far wake region. These SPOD structures reveal a pairing instability that develops downstream, distinct from the near-wake structures captured by \( \mathbf{V}_1 \). A direct comparison with the structures of $\mathbf{V}_1$ at $St=0.156(=2 \times 0.078)$, further supports this interpretation. As the energy and spatial extent of $\mathbf{V}_1$ structures begin to diminish around $x \approx 10$, the structures associated with $\mathbf{V}_2$ begin to grow in intensity. This spatial shift signals an energy transfer from $\mathbf{V}_1$ to $\mathbf{V}_2$, corresponding to the nonlinear pairing of vortices and the emergence of the subharmonic. In contrast to $\mathbf{V}_1$, which represents the instability of individual vortices in the near wake, $\mathbf{V}_2$ reflects a slower-growing instability that dominates the far-wake and emerges from vortex pairing. 

The two unstable CLVs together capture all the distinct but interconnected mechanisms, that is the primary instabilities due to vortex shedding and jet-flapping as well as the secondary instability driven by vortex merging.

\section{Global linear stability analysis of the time-average flow \label{sec:global_LSA}}

In this section we report and discuss results from global linear stability analysis (GLSA) around the time-average flow, and compare with the results of the Lyapunov analysis presented earlier. This comparison will elucidate similarities and differences between the two approaches. 

The starting point for GLSA in the linearization of the Navier-Stokes equations about the time-averaged base flow shown in figure \ref{fig:timeaveraged_baseflowU}. After discretization the linearized system takes the symbolic form,
\begin{equation}
\frac{d \mathbf{u}^\prime}{d t} = A \mathbf{u}^\prime,
\label{eq:LSA_equation}
\end{equation}
where \(\mathbf{u}^\prime(t)\) represents the discrete perturbation velocity vector stored at the cell centers and $A$ is the discrete and constant in time linearized Navier-Stokes operator. The analytic solution after a time window $\Delta t$ is:
\begin{equation}
\mathbf{u}^\prime(\Delta t) = B(\Delta t) \mathbf{u}^\prime(0), \quad B(\Delta t) = e^{A \Delta t}
\end{equation}
where \(B(\Delta t)\) is known as the matrix exponential and represents the evolution operator of \eqref{eq:LSA_equation} over $\Delta t$. The eigenvalues of  $A$ ($\lambda_A$) can be obtained from those of $B$ ($\lambda_B$) from
\begin{equation}
\lambda_{A} = \frac{\ln(\lambda_{B})}{\Delta t},
\end{equation}
The two matrices share the same eigenvectors. 

Given the large size of $A$, direct methods cannot be applied. This necessitates the use of iterative techniques, such as Krylov subspace methods \citep{edwards_krylov_1994, Lehoucq}, that project the high-dimensional system onto a lower-dimensional subspace, where direct methods can be applied. The Krylov subspace \(K\), spanned by snapshots of the flow perturbations \(\mathbf{u}^\prime\), is constructed as follows:
\begin{equation}
K = \text{span} \{\mathbf{u}^\prime(0), \mathbf{u}^\prime(\Delta t) , ..., \mathbf{u}^\prime((m-1)\Delta t)\}
=\text{span} \{\mathbf{u}^\prime(0), B\mathbf{u}^\prime(0),..., B^{m-1}\mathbf{u}^\prime(0) \}
\end{equation}
where \(\mathbf{u}^\prime(0)\) is the (random) initial perturbation field. The Krylov subspace is orthonormalized yielding a unitary basis \(V\) onto which the matrix exponential is projected $B \approx V H V^\top$, where \(H\) is an upper Hessenberg matrix. This results in a small \(m \times m\) eigenvalue problem of the form \(HS = \Sigma S\), with \(\Sigma = \text{diag}(\sigma_1,...,\sigma_m)\) which can be easily solved. More details can be found in \cite{ barkley_three-dimensional_2002, schmid_nonmodal_2007, bagheri_global_2009}.


We use the implicitly restarted Arnoldi method (IRAM), implemented in the software package ARPACK \citep{Lehoucq}. The dimension of the Krylov subspace is set to $m = 20$. The first initial guess was random noise. For the convergence of the first six eigenmodes largest real part  approximately $1200$ iterations are required. The residual was set to \(\|B\phi_j - \sigma_j \phi_j\| < 10^{-4}\) for all eigenmodes. 

\begin{figure}[ht!]
    \centering    
    \includegraphics[width=1\linewidth]{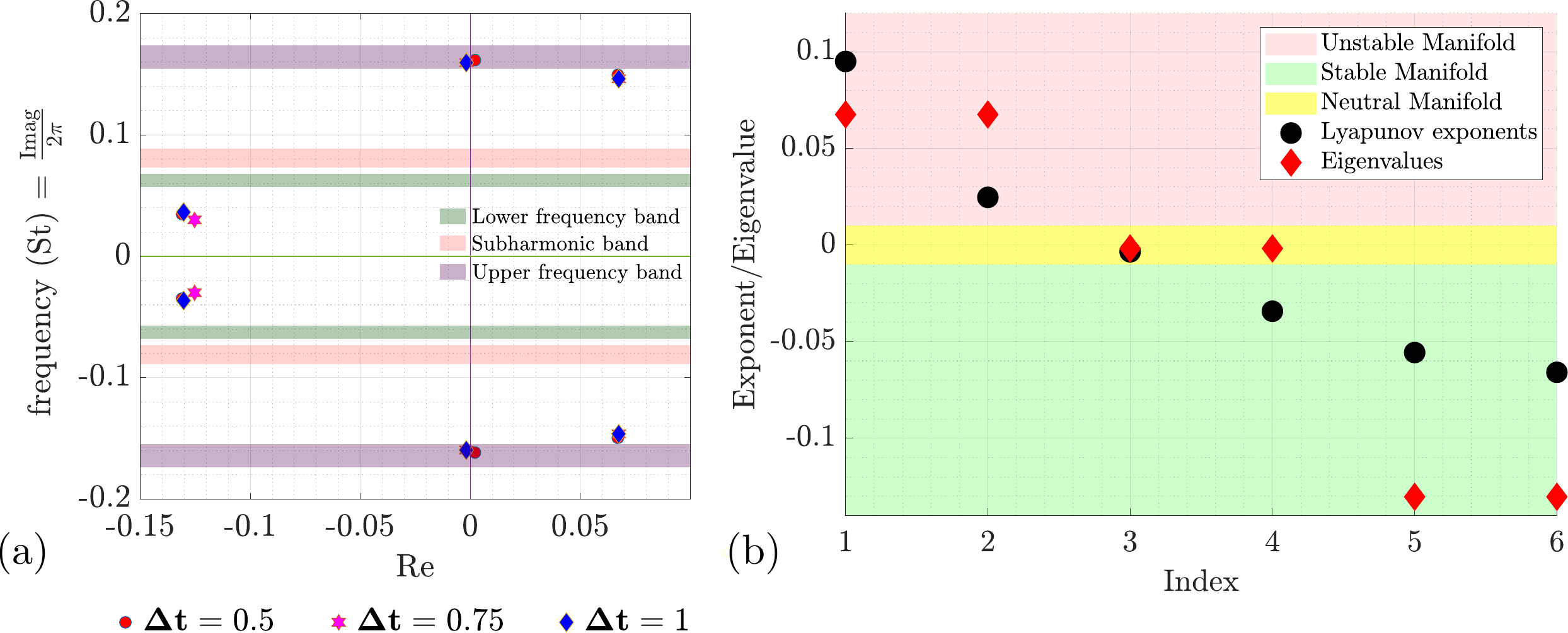}
\caption{Eigenvalues from GLSA of time-averaged flow (left) Comparison of LEs and the real parts of the eigenvalues from GLSA (right).}
    \label{Real_imag_LE_eigval}
\end{figure}

The (complex) eigenvalues are sorted by the real part (that represents the growth rate) and plotted in the left panel of figure \ref{Real_imag_LE_eigval}. Results were obtained for three values $\Delta t = 0.5$, $0.75$ and $1$. To facilitate the interpretation of the results, the imaginary part is converted into frequency. As can be seen, the eigenvalues are very close to each other, confirming the accuracy of the GLSA. The most unstable eigenmode pair oscillates at $St = 0.149$, the neutral eigenmode pair oscillates at a slightly higher frequency of $St = 0.161$, and the stable pair oscillates at $St = 0.035$. Notice that the frequency of the neutral pair captures the shedding frequency very well and thus is within the upper frequency band. Similar observations have been made in the past. For example \cite{Barkley_2006} applied global linear stability analysis to the time-averaged wake of a circular cylinder ($Re \in [46,180]$) and demonstrated that the leading eigenmode is neutrally stable while its frequency exactly matches the measured Strouhal number. The present case is however slightly different because we have both an unstable and a neutral eigenmode (in the single cylinder case there was only a neural mode). 

The other observation is that the lower frequency band and the subharmonic are not captured by the GLSA. This is consistent with the results of \cite{Carini_Giannetti_Auteri_2014} who found that the jet-flapping (that oscillates at the low frequency) arises as an instability of the periodic in-phase synchronised shedding and can be revealed only with Floquet stability analysis; standard GLSA therefore cannot detect it.


The right panel of figure \ref{Real_imag_LE_eigval} collates the real part of the GLSA eigenvalues and the LEs in a single plot. It is interesting to notice that the LEs, which quantify the growth rates of perturbations around an unsteady base flow, are of the same order of magnitude as the growth rates of perturbations around the time-averaged flow. However, while an unstable eigenvalue around a time-average flow is not physically meaningful, the LEs are physically meaningful invariant quantities of the flow.

\begin{figure}[ht!]
    \centering
    \includegraphics[width=1\linewidth]{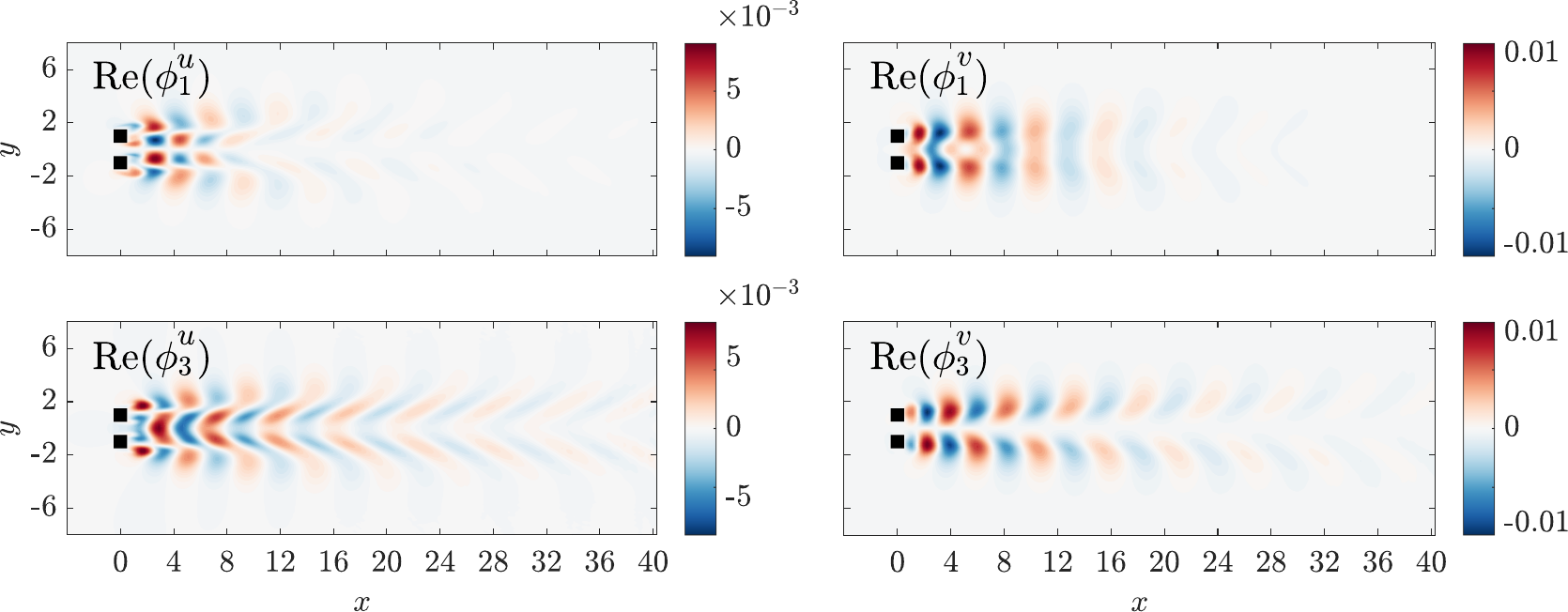}
    \caption{Contour plots of streamwise velocity (left column) and cross-stream velocity (right column)  of the real parts of the unstable eigenvector  $\phi_1$ (top row) and the neutral eigenvector $\phi_3$ (bottom row).}
    \label{fig:eigenvectors_phi1_phi3}
\end{figure}

Contours of the real part of the streamwise and cross-stream components are shown in figure \ref{fig:eigenvectors_phi1_phi3} for the unstable (top row) and neutral (bottom row) eigenvectors. In the near wake behind each cylinder, the cross-stream component of the unstable eigenvector, $\phi_1^v$, has spatial footprint not only behind each cylinder but in the gap centerline  as well, with structures of alternating sign. Both $u$ and $v$ structures are symmetric, indicating in-phase activity. The observed pattern however does not correspond to any of the patterns observed earlier in the SPOD analysis of the flow or the CLVs. 
On the other hand, the component $\phi_3^v$ of the neutral eigenvector displays structures of alternating sign behind each cylinder which are 
similar to the dominant SPOD modes of the flow (see figure \ref{fig:SPOD_nonlinear}(b)), and that of $\mathbf{V_1}$ (see $\Phi_{1r}^{0.156}$ in figure \ref{SPOD_CLV1_contours}). 
The structures however identified through GLSA exhibit a mild growth in the cross-stream direction as they propagate downstream. In contrast, the structures observed in the Lyapunov stability analysis, specifically the SPOD modes of $\mathbf{V_2}$, exhibit a stronger (and much more physically realistic) lateral expansion as seen in figure \ref{SPOD_CLV2_contours}.




The above observations underscore the limitations of GLSA. While it can identify the shedding mode and approximate well its  frequency, it may fail to capture the true unstable nature of some directions that the Lyapunov analysis reveals.

\section{Conclusion}\label{sec:conclusions}

We consider the irregular two-dimensional flow around two square cylinders at $Re = 200$ from a dynamical systems perspective. Combining inspection of vorticity dynamics as the flow evolves, Lyapunov stability analysis, hyperbolicity assessment, and global linear stability analysis, we have uncovered new insights into the instability mechanisms of the flow.

Lyapunov stability analysis confirms that the flow is chaotic with two positive LEs. The CLVs associated with these LEs characterize the unstable directions in the tangent space and enable visualization of the instability regions of the flow. The present study has considered only two-dimensional perturbations (i.e.\ perturbations that have zero wavenumber in the spanwise direction), but other wavenumbers can be easily considered. Time-averaged contours reveal that the most unstable CLV is concentrated in the near-wake region, whereas the second unstable CLV is prominent further downstream. The stable CLVs are  active in the far wake.  Spectral Proper Orthogonal Decomposition (SPOD) of the leading CLVs reveals the dominant coherent structures and their oscillation frequencies. For the first CLV, the two most energetic frequencies identified by SPOD fall within the dominant upper and lower frequency bands of the $C_D$ spectrum. In contrast, the second unstable CLV contributes primarily to dynamics in the mid-wake and is associated with the process of vortex merging that leads to a subharmonic instability. The angle between individual CLVs were computed to assess the hyperbolicity of the flow. It was found that CLVs in the neutral and stable subspaces can become nearly tangent.

GLSA applied to the time-averaged flow yields a neutral eigenvector pair whose spatial structures closely resemble those of the leading SPOD mode of the most unstable CLV. Moreover, the oscillation frequency matches very closely with that of the non-linear flow. However, while GLSA identifies this mode as neutral, the Lyapunov-based analysis reveals that it is growing. Furthermore the GLSA cannot capture the jet-flapping and the subharmonic instabilities. These discrepancies highlight the limitations of global LSA of the time-average flow and demonstrate that Lyapunov analysis can provide a much more realistic picture of the underlying instability dynamics.

We close the paper with some thoughts on future directions that this research can open. Recall that for periodic flows the CLVs are identical to the Floquet modes. In this sense, Lyapunov stability analysis is a generalisation of Floquet analysis to general irregular flows (in the same way that Floquet was a generalisation of the standard GLSA to periodic flows). Therefore Lyapunov stability offers a unifying framework that encompasses both Floquet and GLSA. The fact that for Lyapunov stability analysis there is no restriction on the underlying base flow allows us to take a seamless approach to stability after a  threshold has been crossed. For example, using  Floquet, one has to use special techniques to obtain a periodic base flow for values of the characteristic parameter  beyond the critical, see \cite{Carini_Giannetti_Auteri_2014}. With Lyapunov stability analysis, this is no longer necessary. Another very promising direction of research is the extension of structural sensitivity analysis  \citep{Giannetti_Luchini_2007} to chaotic flows. However, this extension is not straightforward because it requires integration of the adjoint equations backwards in time. Due to the chaotic nature of the flow however the adjoint variables exponentially diverge. One however can use the least-squares shadowing technique \citep{Ni2017, Kantarakias_Papadakis_2023b, Kantarakias_Papadakis_PRE_2024} to compute the adjoint variables  and from those the structural sensitivity. We hope that these are fruitful research directions that will uncover a wealth of new physics in complex flows.

\begin{appendix}

\section{CLV calculation methodology \label{CLVcalcMethodappendix}} 
Below we summarise the algorithm of \cite{ginelli_covariant_2013} to obtain CLVs. We use the notation
\begin{equation}
\mathbf{u}_m=\mathbf{f}^{(m)}\left(\mathbf{u}_0\right), \quad 
\mathbf{u}_{m+n}=\mathbf{f}^{(m)} \left(\mathbf{f}^{(n)}\left(\mathbf{u}_0\right)\right)=
\mathbf{f}^{(m)}\left(\mathbf{u}_n\right)=\mathbf{f}^{(n)}\left(\mathbf{u}_m\right)
\end{equation}
\noindent where $\mathbf{u}$ is the state vector of size $2N$, and $\mathbf{f}$ is the evolution operator that represents the dynamics of the system. The superscripts $m$ and $n$ denote different time instants. We assume that the map $\mathbf{f}$ is invertible, that is $\mathbf{f}^{(-m)}(\mathbf{u_m}) = \mathbf{u_0}$. 

The Jacobian at a state $\mathbf{u}_n$ can be written as,
\begin{equation}
\mathbf{J}\left(\mathbf{u}_n\right)=\frac{\partial \mathbf{f}\left(\mathbf{u}_n\right)}{\partial \mathbf{u}_n},
\end{equation}

Based on \eqref{kol_lin1_eq}, the tangent space evolution operator that takes the perturbation field from time instant $n$ to time instant $k+n$  is the product,
\begin{equation}
\mathbf{M}_{k, n}=\prod_{i=n}^{k+n-1} \mathbf{J}\left(\mathbf{u}_i\right),
\end{equation}
Assume now the orthonormal Gram-Schmidt (GS) vectors at $n$, $\mathbf{G}_n = \left(\mathbf{g}_n^{(1)}\left|\mathbf{g}_n^{(2)}\right| \ldots \mid \mathbf{g}_n^{(2N)}\right)$. The operation  
\begin{equation}
\mathbf{M}_{k, n} \mathbf{G}_n=\tilde{\mathbf{G}}_{k+n}
\label{eq:propagation_of_GS_vectors}
\end{equation}
propagates these vectors to $k+n$, i.e.\ $\tilde{\mathbf{G}}_{k+n}$ is the evolved tangent space at time $k+n$. Note that the columns of $\tilde{\mathbf{G}}_{k+n}$ are no longer orthonormal. Applying Gram-Schmidt QR decomposition to $\tilde{\mathbf{G}}_{k+n}$ gives,
\begin{equation}
\mathbf{M}_{k, n} \mathbf{G}_n=\mathbf{G}_{k+n} \mathbf{R}_{k, n},
\label{growing_Gn_eq}
\end{equation}

Let $\mathbf{V}_n  = \left(\mathbf{v}_n^{(1)}\left|\mathbf{v}_n^{(2)}\right| \ldots \mid \mathbf{v}_n^{(2N)}\right)$ denote the 2N CLVs at time $n$. The $i$-th CLV, $\mathbf{v}_n^{(i)}$, can be expanded in terms of the GS basis vectors $\mathbf{g}_n^{(j)}$ as,
\begin{equation}
\mathbf{v}_n^{(i)}=\sum_{j=1}^i c_n^{(j, i)} \mathbf{g}_n^{(j)}
\label{eq:clv_exp_coeff}
\end{equation}
\noindent where $c_n^{(j, i)}$ is known as a CLV expansion coefficient. Note that $\mathbf{v}_n^{(i)}$ depends only on the GS vectors $j=1\dots i$. This means that the first CLV coincides with the first vector of the GS basis by construction. Moreover, since CLVs are have unit norm and $\mathbf{g}_n^{(j)}$ are orthonormal, the following applies,
\begin{equation}
    \sum_{j=1}^i\left(c_n^{(j, i)}\right)^2=1,
    \label{cnnormaleq}
\end{equation}
for all $i$.

Equation \ref{eq:clv_exp_coeff} in matrix notation is written as,
\begin{equation}
\mathbf{V}_n = \mathbf{G}_n \mathbf{C}_n 
\label{CLVeq1}
\end{equation}
where $\mathbf{C}_n$ is an upper-triangular matrix that contains the CLV expansion coefficients, $\left[\mathbf{C}_n\right]_{j, i}=c_n^{(j, i)}$ for $j \leqslant i$. 

Applying the propagation operator $\mathbf{M}_{k, n}$ to $\mathbf{V}_n$ we get
\begin{equation}
\mathbf{M}_{k, n} \mathbf{V}_n=\mathbf{\tilde{V}}_{n+k}
\label{CLVeq2}
\end{equation}
but $\mathbf{\tilde{V}}_{n+k}$ is no longer orthonormal. Using QR decomposition, we can write
\begin{equation}
\mathbf{\tilde{V}}_{n+k}=\mathbf{V}_{n+k} \mathbf{D}_{k, n}
\label{CLVeq3}
\end{equation}
where  $\mathbf{D}_{k, n}$ is a diagonal matrix composed of the local growth factors $\gamma_{k, n}^{(i)}=\left\|\mathbf{M}_{k, n} \mathbf{v}_n^{(i)}\right\|$, i.e.\ $\left[\mathbf{D}_{k, n}\right]_{i, j}=\delta_{i, j} \gamma_{k, n}^{(i)}$. For finite $k$, the logarithms of the growth factors are the finite-time Lyapunov exponents (FTLE) and their time-average obviously coincides with the LEs. 

Combining \eqref{CLVeq2} and \eqref{CLVeq3}, we obtain the following equation that governs for evolution of the CLVs from $n$ to $k+n$, 
\begin{equation}
\mathbf{M}_{k, n} \mathbf{V}_n=\mathbf{V}_{n+k} \mathbf{D}_{k, n}
\label{CLVeq4}
\end{equation}
 
Substituting \eqref{CLVeq1} into \eqref{CLVeq4} we obtain,
\begin{equation}
\mathbf{M}_{k, n} 
\mathbf{G}_n \mathbf{C}_n =
 \mathbf{G}_{n+k} \mathbf{C}_{n+k}  \mathbf{D}_{k, n}
\label{eq_penultimate1}
\end{equation}

Using \eqref{growing_Gn_eq} we can write, 
\begin{equation}
  \mathbf{M}_{k, n} 
\mathbf{G}_n \mathbf{C}_n = \mathbf{G}_{n+k} \mathbf{R}_{k, n} \mathbf{C}_n
\label{eq_penultimate2}
\end{equation}

Equating the right hand sides of \eqref{eq_penultimate1} and \eqref{eq_penultimate2}, we obtain 
\begin{equation}
 \mathbf{G}_{n+k} \mathbf{C}_{n+k}  \mathbf{D}_{k, n}=\mathbf{G}_{n+k} \mathbf{R}_{k, n} \mathbf{C}_n
\end{equation}
and solving for $\mathbf{C}_n$ we get,
\begin{equation}
\mathbf{C}_n=\mathbf{R}_{k, n}^{-1} \mathbf{C}_{n+k} \mathbf{D}_{k, n},
\end{equation}
which forms the backward evolution equation for the upper-triangular matrix $\mathbf{C}_n$. Here $\mathbf{R}_{k, n}^{-1}$ are the inverted upper-triangular matrices which are by-products of the forward evolution of the GS vectors, see \eqref{growing_Gn_eq}.

Since $\mathbf{C}_n$ has to be normalized column by column, see \eqref{cnnormaleq}, the diagonal matrix $\mathbf{D}_{k, n}$ can be neglected, thus 
\begin{equation}
    \mathbf{C}_n=\mathbf{R}_{k, n}^{-1} \mathbf{C}_{n+k}.
    \label{CLVbackeq}
\end{equation}
\cite{ginelli_covariant_2013} proved that (almost) any non-singular upper-triangular matrix when evolved backwards according to \eqref{CLVbackeq} can generate the appropriate CLV expansion coefficients. 

The computation of CLVs requires the GS vectors and CLV expansion coefficients. Once these are calculated, the CLVs can be obtained from \eqref{CLVeq1} in the middle time window where the GS vectors and the expansion coefficients have converged.

\end{appendix}

\section*{Acknowledgements}{The authors would like to thank Dr Liang Fang for numerous discussions on the topics of the paper. High-performance computing resources were provided by Imperial College London.}

\section*{Funding}{G. Papadakis is supported by EPSRC (grant EP/W001748/1) and NERC (grant NE/Z503861/1)}

\section*{Declaration of interests}{The authors report no conflict of interest.}

\section*{Data availability statement} 
{The numerical codes used for the calculation of LEs and CLVs are released as open‑source software at \url{https://github.com/sidsahuCFD/lyapunov-stability-fluid-dynamics}.}


\section*{Author ORCIDs}{Sidhartha Sahu \url{https://orcid.org/0009-0004-6304-9036}, George Papadakis \url{https://orcid.org/0000-0003-0594-3107}.}

\section*{Author contributions}{S. Sahu conducted the simulations, data analysis, and drafted the manuscript. G. Papadakis obtained the funding, supervised the project, contributed to the analysis of the results, and revised the manuscript. Both authors approved the final version of the manuscript.}





\bibliographystyle{plainnat}
\bibliography{jfm.bib}
\newpage \clearpage

\end{document}